\documentclass[usegraphicx]{aa}

\usepackage{natbib}
\bibpunct{(}{)}{;}{a}{}{,} % to follow the A&A style
\usepackage{subfigure}
\usepackage{longtable}
\usepackage{graphicx}
\usepackage{graphics}
\usepackage{keyval}
\usepackage{trig}
\usepackage[american]{babel}
\usepackage{url}
\usepackage{color}
\usepackage{amsmath}
\usepackage{bm}
\usepackage{longtable}

\def\beq{\begin{equation}}
\def\eeq{\end{equation}}
\def\bey{\begin{eqnarray}}
\def\eey{\end{eqnarray}}

\def\lsim{\mathrel{\raise.3ex\hbox{$<$\kern-.75em\lower1ex\hbox{$\sim$}}}}
\def\gsim{\mathrel{\raise.3ex\hbox{$  $\kern-.75em\lower1ex\hbox{$\sim$}}}}

\def\a0{A_{\mathrm{0}}}

\def\teff{T_\mathrm{eff}}
\def\fehzw{\rm [Fe/H]_{ZW}}
\def\fehuves{\rm [Fe/H]_{UVES}}

\def\apjl{ApJ}

\def\aap{A\&A}      % Astronomy and Astrophysics (Letters indicated by number)
\def\apjs{ApJS}       % Astrophysical Journal Supplement Series (the)
\newcommand\Eq[1]{Eq.~(\ref{#1})}
\newcommand\Fig[1]{Fig.~\ref{#1}}
\newcommand\Tab[1]{Table~\ref{#1}}
\newcommand\Sec[1]{Sec.~\ref{#1}}
\begin{document}
   \title{Constraining the parameters of globular cluster NGC 1904 from its variable star population}

   \author{N. Kains
          \inst{1}\fnmsep\thanks{nkains@eso.org}
          \and
          D. M. Bramich\inst{1}
          \and
          R. Figuera Jaimes\inst{1, 2}
          \and
          A. Arellano Ferro\inst{3}
          \and
          S. Giridhar\inst{4}
          \and
          K. Kuppuswamy\inst{4}
          }

   \institute{European Southern Observatory, Karl-Schwarzschild Stra\ss e 2, 85748 Garching bei M\"{u}nchen, Germany\\
              \email{nkains@eso.org}
         \and
         SUPA School of Physics \& Astronomy, University of St Andrews, North Haugh, St Andrews, KY16 9SS, United Kingdom \\
         \and
             Instituto de Astronom\'{i}a, Universidad Nacional Aut\'{o}noma de Mexico\\
      \and
      	Indian Institute of Astrophysics, Koramangala 560034, Bangalore, India	
             }

   \date{Received ... ; accepted ...}

% \abstract{}{}{}{}{} 
% 5 {} token are mandatory

  \abstract
  % context heading (optional)
  % {} leave it empty if necessary  
   {}
  % aims heading (mandatory)
   {We present the analysis of 11 nights of $V$ and $I$ time-series observations of the globular cluster NGC 1904 (M 79). Using this we searched for variable stars in this cluster and attempted to refine the periods of known variables, making use of a time baseline spanning almost 8 years. We use our data to derive the metallicity and distance of NGC 1904.}
  % methods heading (mandatory)
   {We used difference imaging to reduce our data to obtain high-precision light curves of variable stars. We then estimated the cluster parameters by performing a Fourier decomposition of the light curves of RR Lyrae stars for which a good period estimate was possible. %We also derive an estimate for the age of the cluster by fitting theoretical isochrones to our colour-magnitude diagram (CMD).
   }
  % results heading (mandatory)
   {Out of 13 stars previously classified as variables, we confirm that 10 are bona fide variables. We cannot detect variability in one other within the precision of our data, while there are two which are saturated in our data frames, but we do not find sufficient evidence in the literature to confirm their variability. We also detect a new RR Lyrae variable, giving a total number of confirmed variable stars in NGC 1904 of 11. Using the Fourier parameters, we find a cluster metallicity $\fehzw=-1.63 \pm 0.14$, or $\fehuves=-1.57 \pm 0.18$, and a distance of $13.3 \pm 0.4$ kpc (using RR0 variables) or 12.9 kpc (using the one RR1 variable in our sample for which Fourier decomposition was possible).}
  % conclusions heading (optional), leave it empty if necessary 
   {}


   \maketitle
%
%________________________________________________________________

% =====================================================
\section{Introduction}\label{sec:intro}
% ====================================================

The $\Lambda$CDM cosmological paradigm predicts that the Milky Way formed through the merger of small galaxies, following a hierarchical process \citep{diemand07}. As potential survivors of these processes, globular clusters are therefore important probes into the formation and early evolution of the Milky Way. Furthermore, they also allow us to delve into the structure of the Galaxy, for example in constraining the shape of the Galactic Halo \citep{lux12}, and to study stellar populations.

There are various methods in use to determine a cluster's properties, including analysis of its colour-magnitude diagram, or obtaining spectroscopy of giant stars in the cluster to determine their characteristics. Another way to determine the metallicity and distance of a cluster, as well as to obtain a lower limit on its age, is to study the population of RR Lyrae variables that reside within it. By obtaining photometry with sufficient time resolution, the light curve of the RR Lyrae stars can be analysed with Fourier decomposition, yielding parameters that can then be used to determine many of the stars' intrinsic properties, thanks to various empirical or theoretical relations derived from collected observations or theoretical models \citep{simon93, jurcsik96, jurcsik98, kovacs98, morgan07}. These can also then be used as proxies for the properties of the stars' host cluster.

In this paper we use this method to determine the properties of the RR Lyrae stars in NGC 1904 (M 79; $\alpha=05^h24^m10^s, \delta=-24^\circ31'27''$ at J2000.0), a globular cluster at $\sim 13$kpc, with metallicity [Fe/H]$\sim -1.6$, and a prominent blue horizontal branch. We use Fourier decomposition to constrain the properties of the cluster itself, following previous papers on other clusters \citep{arellano04, lazaro06, arellano08a, arellano08b, arellano10, bramich11, arellano11}. This cluster has been poorly studied with no major published long-baseline time-series CCD observations, although it contains 13 stars classified as variables, including 5 detected in a recent study by \cite{amigo11}. Here we present an analysis using data spanning almost 8 years, allowing us to improve the accuracy of the RR Lyrae periods and to detect variations in period or amplitude that might be indicative of the presence of the Blazhko effect.

Although it has been suggested NGC 1904 belonged to the Canis Major (CMa) dwarf galaxy \citep{martin04}, other studies (e.g. \citealt{mateu09}) have questioned this due to the lack of blue horizontal branch stars in the CMa colour-magnitude diagram (CMD) compared to this and other clusters also potentially associated with the CMa overdensity, such as NGC 1851, NGC 2298 and NGC 2808. Further study of these clusters is therefore crucial to improving our understanding of the structure of the Galactic disc, as \cite{martin04} also suggested that CMa is made up of a mixture of thin and thick disc and spiral arm populations of the Milky Way.

In \Sec{sec:observations}, we detail our observations, and in \Sec{sec:variables}, we discuss variable objects in NGC 1904 both in the literature and those which we detect in our data. In \Sec{sec:fourdec} we perform Fourier decomposition of the light curves of some of the variables to determine their properties. We then use these in \Sec{sec:clusterprop} to calculate values for the cluster parameters. We place these values into context in \Sec{sec:mvfeh} by comparing them to those found for other clusters in the literature. Finally, we also discuss briefly the peculiar spatial distribution of the RR Lyrae variables that we find in NGC 1904.

\section{Observations and reductions}\label{sec:observations}

\subsection{Observations}
We obtained Johnson $V$-band data of NGC 1904 using the 2.0m telescope at the Indian Astrophysical Observatory in Hanle, India, on 11 nights spanning from April 2004 to March 2012, a baseline of almost 8 years. The cluster was also observed in the $I$ band, except in 2004 when no $I$ data was taken. This resulted in 147 $V$-band and 109 $I$-band images. The observations are summarised in \Tab{tab:observations}.

% =====================================================
\begin{table}
\centering
\begin{tabular}{@{}lcccccc@{}}
\hline
Date     & $N_{V}$ & $t_{V}$ (s)  & $N_{I}$ & $t_{I}$ (s) \\
\hline
20041004 & 7      &  60-200            & 0       & --- \\
20041005 & 3      &    100                  & 0       & --- \\
20090108 & 16      &   150                & 14       & 120 \\
20100123 & 19      &   80-150         & 21       & 40-120 \\
20100221 &  16      & 100-200      & 14       & 50-100 \\
20100307 &  15      & 90-120        & 13       & 45-55 \\
20110312 &  16      &   120-140       & 0       & --- \\
20110313 & 10      &   80-150           & 0       & --- \\
20111104 &  30      &   90-300           & 32       & 25-120 \\
20120301 & 10      & 150-180            & 10        & 50-65 \\
20120302 &  5      &   100-120           & 5       & 45 \\
\hline
Total:   &  147    &                    &  109      &     \\
\hline
\end{tabular}
\caption{The distribution of observations of NGC 1904 for each filter, where the columns $N_{V}$ and $N_{I}$ represent
         the number of images taken for the filters $V$ and $I$, respectively.
         We also provide the exposure time, or range of exposure times, employed during each night for each filter
         in the columns $t_{V}$ and $t_{I}$.\label{tab:observations}}
\end{table}
% =====================================================

The CCD used was a Thompson 2048 $\times$ 2048 pixel, with a field of view of 10.1 $\times$ 10.1 arcmin$^2$, giving a pixel scale of 0.296 arcsec per pixel. Given the distance to the cluster we derive later in this paper, this corresponds rougly to an area of 40 pc $\times$ 40 pc centred on the core of the cluster.

% =====================================================
\begin{figure}
  \centering
\includegraphics[width=8cm, angle=0]{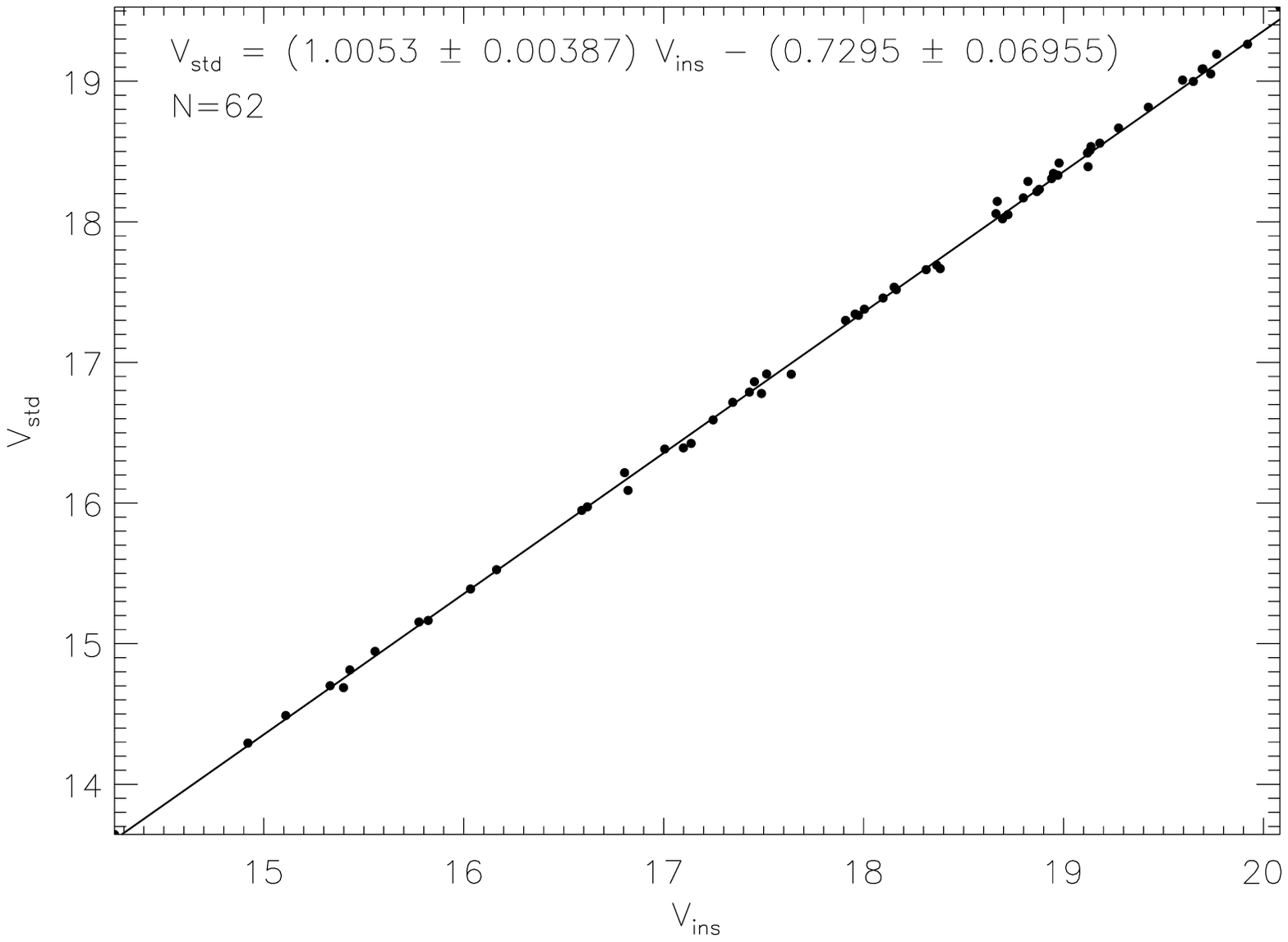}
\includegraphics[width=8cm, angle=0]{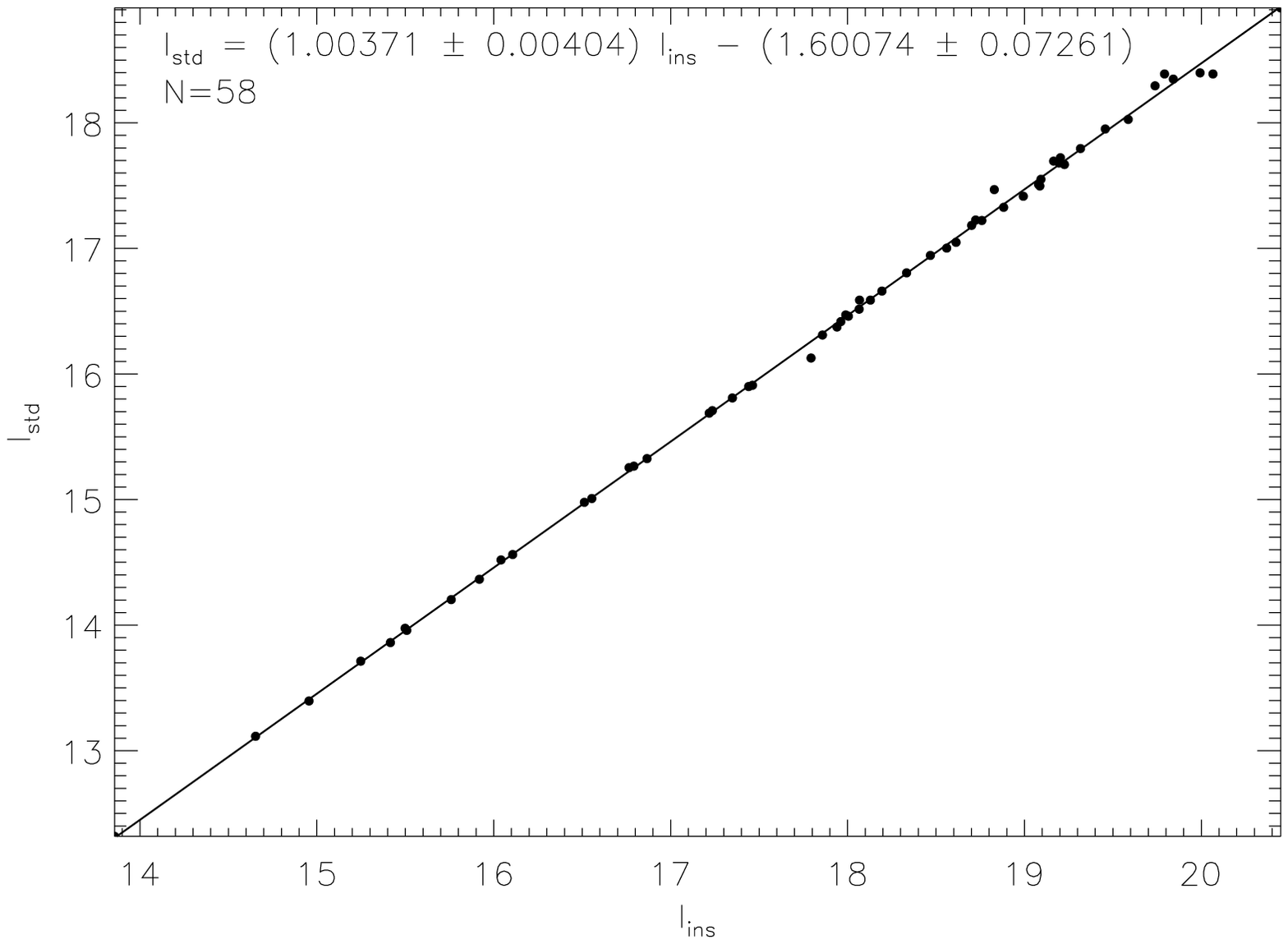}

\caption{Linear regression fits to the relation between instrumental and Johnson-Kron-Cousins magnitudes for the $V$ (top) and $I$ (bottom) bands. \label{fig:transf}}

\end{figure}
% =====================================================

\subsection{Difference Image Analysis}\label{sec:dia}

We reduced our observations using the {\tt DanDIA}\footnote{{\tt DanDIA} is built from the {\tt DanIDL} library of {\tt IDL} routines available at {\tt http://www.danidl.co.uk}} pipeline (\citealt{bramich08, bramich12b}) to obtain high-precision photometry of sources in the images of NGC 1904, as was done in previous globular cluster studies (e.g. \citealt{bramich12}, see that paper for a detailed description of the software). We produce a stacked reference image in each band by selecting the best-seeing images (within 10\% of the best-seeing value) and taking care to minimise the number of saturated stars. Our resulting reference image in the $V$ filter consists of a single image with an exposure time of 100 seconds, and full width half-maximum (FWHM) of the point spread function (PSF) of 5.37 pixels ($\sim 1.5''$). In the $I$ band, the reference image is made from 3 stacked images with a total exposure time of 150 seconds and a PSF FWHM of 4.79 pixels ($\sim 1.4''$).

Because of the way difference imaging works, the measured reference flux of a star on the reference image might be systematically too large due to contamination from other nearby objects (blending). Since non-variable sources are fully subtracted on the difference images, however, this problem does not occur with the difference images. Hence variable sources with overestimated reference fluxes will have an underestimated variation amplitude, although crucially the shape of the light curve is unaffected. Furthermore, some of the targets which are in the vicinity of saturated stars or are saturated themselves will have less precise or non-existent photometry. Although we tried to minimise the number of saturated stars in our reference images, some remained and this affected the photometry of some of the objects in this cluster.

% =====================================================
\begin{table*}
\begin{center}
  \begin{tabular}{ccccccccccc}

     \hline
    \#		&Filter	&HJD 	&$M_{\rm std}$	  & $m_{\rm ins}$ 	&$\sigma_m$		&$f_{\rm ref}$ 	&$\sigma_{\rm ref}$	& $f_{\rm diff}$ &$\sigma_{\rm diff}$ &$p$  \\
	&	&($d$)	&(mag)	&(mag)	&(mag)	&(ADU s$^{-1}$)	&(ADU s$^{-1}$)	&(ADU s$^{-1}$)	&(ADU s$^{-1}$)	&	\\
  \hline  
   
V3 &V & 2453283.39244 &    16.235 &    16.875 &     0.005 &  2531.207 &     6.272 &  -740.195 &     7.926 &    0.9831 \\
V3 &V & 2453283.39990 &    16.237 &    16.877 &     0.003 &  2531.207 &     6.272 &  -748.391 &     4.372 &    0.9900 \\
\vdots &\vdots&\vdots&\vdots&\vdots&\vdots&\vdots&\vdots&\vdots&\vdots&\vdots \\
V3 &I & 2454840.11764 &    15.555 &    17.093 &     0.005 &  1206.297 &     6.739 &   380.674 &     9.788 &    1.5286 \\
V3 &I & 2454840.12374 &    15.566 &    17.104 &     0.006 &  1206.297 &     6.739 &   358.597 &    11.513 &    1.5294 \\
\vdots &\vdots&\vdots&\vdots&\vdots&\vdots&\vdots&\vdots&\vdots&\vdots&\vdots \\

\hline \hline
  \end{tabular}
  \caption{Format for our time-series photometry, for all confirmed variables in our field of view. The standard $M_{\rm std}$ and instrumental $m_{\rm ins}$ magnitudes listed in column 4 and 5 respectively correspond to the variable star, filter and epoch of mid-exposure listed in columns 1-3, respectively. The uncertainty on $m_{\rm ins}$ is listed in column 6, which also corresponds to the uncertainty on $M_{\rm std}$. For completeness, we also list the reference flux $f_{\rm ref}$ and the differential flux $f_{\rm diff}$ (columns 7 and 9 respectively), along with their uncertainties (columns 8 and 10), as well as the photometric scale factor $p$. Definitions for these quantities can be found in e.g. \cite{bramich11}, Eq. 2-3. This is a representative extract from the full table, which is available with the electronic version of the article (see Supporting Information). \label{tab:onlinedata}}
  \end{center}
\end{table*}
% =====================================================

\subsection{Photometric Calibrations} 

Instrumental magnitudes were converted to standard Johnson-Kron-Cousins magnitudes by fitting a linear relation between the known magnitudes of standard stars in NGC 1904 \citep{stetson00} and the light curve mean magnitudes. These relations, plotted in \Fig{fig:transf}, were then used to convert all instrumental $V-$ and $I-$ band light curves to standard magnitudes. We note that the standard stars we use cover the full magnitude and colour ranges of our CMD, justifying our choice not to include a colour term in our calibration relations. Calibrated data for all variables is available with the electronic version of this paper, in the format given in \Tab{tab:onlinedata}.

\subsection{Astrometry}

We used the online tool {\tt astrometry.net} \citep{lang09} to obtain an astrometric fit to our $V-$band reference image, and used this to calculate the J2000.0 coordinates of all objects we discuss in this paper. This tool uses the USNO-B catalog of astrometric standards to find an astrometric fit to images. The coordinates are those of the epoch of our $V-$band reference image, which was taken at HJD=2453284.41 d. We estimate that the astrometric fit is accurate to within $\sim 0.1''$ in $\alpha$ and $\sim 0.3''$ in $\delta$.

\section{Variables in NGC 1904}\label{sec:variables}

The first five variables (V1-5) in this cluster were reported by \cite{bailey1902}, examining plates taken at the Harvard College station in Arequipa, Peru. Of these, V1 has not been confirmed as variable by subsequent studies, while V2 is classified as a semi-regular variable by \cite{rosino52}. One more variable (V6) was then identified by \cite{rosino52} from observations carried out between 1948 and 1951 with the 24-inch reflector of Lojano; these also provided period estimates for the new variable as well as two of Bailey's original variables (V3-4). Rosino also corrected the coordinates for V3 and V4, and commented that V5 in Bailey's original study was not detectable in the new observations due to how crowded his photographic plates were.

Two more variables (V7-8) were reported by \cite{sawyerhogg73}, although the online catalogue of \cite{clement01} notes that these are unlikely to be RR Lyrae variables, as they should have been detected as such in the subsequent work of \cite{amigo11}. That study analysed $B$-filter observations taken in 2001 at the Danish 1.54m telescope in La Silla using image subtraction, and identified an additional five new RR Lyrae variables in NGC 1904, V9-V13. They also estimated the periods for these, as well as the periods of V3, V4, V5 (which they mistakenly referred to as ``NV6") and V6.

This amounts to a total of 13 stars classified as variables in this cluster, 9 of which have had their period estimated, and 3 of which are suspected variables. Only the most recent time-series study was carried out using CCD cameras \citep{amigo11}, and no work employing multi-band time series observations for this cluster with CCD cameras has been published so far. 

In this paper we use the notation introduced by \cite{alcock00}, referring to fundamental-mode pulsation RR Lyrae stars as RR0 and to first-overtone RR Lyrae pulsators as RR1, rather than RR$ab$ and RR$c$, respectively.

\subsection{Stars that do not show variablity}

V1 did not show signs of variability in our data, to within the limit set by its $V$-band root mean square (rms) scatter of 0.015 mag and $I$-band rms of 0.044 mag. This is in agreement with the finding of \cite{rosino52}, which mentions that V1 might not be variable.

\subsection{Variable candidates without light curves}

V2 and V8 are both saturated in our reference images, and we are therefore unable to present light curves for them. To the best of our knowledge, there is no published light curve for either of these two objects. V8 was only reported as variable in a private communication by Tsoo Yu-Hua to H. Sawyer Hogg (see online catalogue of \citealt{clement01}), and was not detected by \cite{amigo11}, which means it is unlikely to be an RR Lyrae variable. V2 is reported by \cite{rosino52} as a semi-regular or irregular variable, with no period estimate. We suggest that more time-series observations are needed to confirm the true variability of V2 and V8 and we believe that they should remain variable candidates.

\subsection{Detection of known variables}

% =====================================================
\begin{figure*}
  \centering
\includegraphics[width=13cm, angle=0]{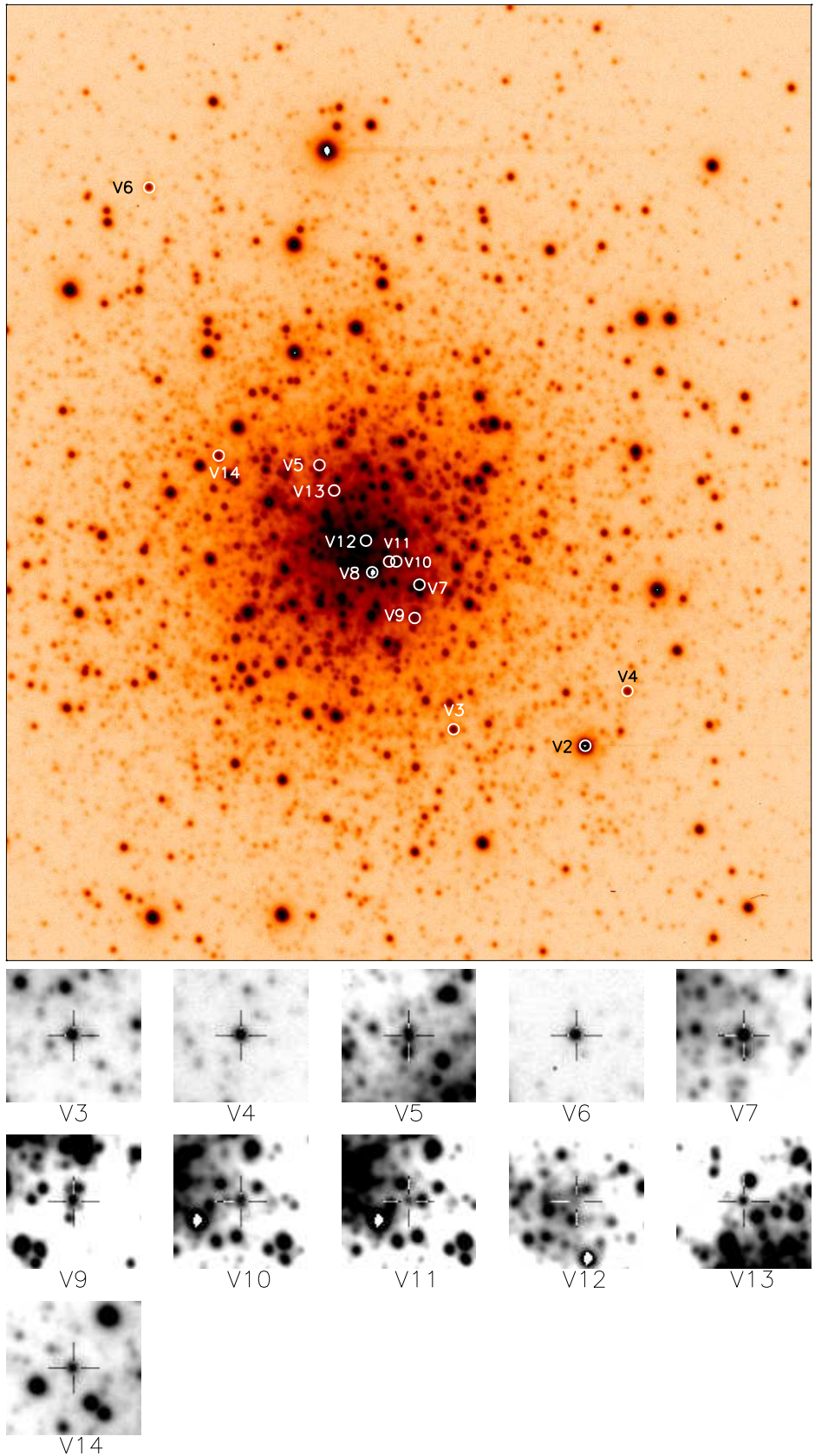}

\caption{Finding chart for the confirmed variable objects in NGC 1904, using our $V$ reference image. North is up and East is to the right. The image size is 4.44 $\times$ 5.18 arcmin$^2$, while each stamp size is 23.7 $\times$ 23.7 arcsec$^2$. White circles and labels indicate the locations of the variables, and each of the variables we detect in our data is shown in an individual stamp, its location marked by a cross-hair. V2 and V8 are saturated in our reference image and therefore we do not present light curves for them, but their locations are shown on this finding chart using the coordinates of \cite{samus09}. The tidal radius $r_t\sim 500''$ \citep{lanzoni07} lies outside of this plot. \label{fig:fchart}}

\end{figure*}
% =====================================================

% =====================================================
\begin{table*}
\begin{center}
  \begin{tabular}{cccccccccc}

     \hline
    \#		 			&Epoch			& $P$ 	&$P$ (Amigo		&$P$  (Rosino 		&$<V>$	&$<I>$	& Amplitude	&Amplitude	& Type \\
    				&(HJD-2450000)	&($d$)	&et al. 2011) ($d$)	& 1952) ($d$)		 	&		&		& ($V$ mag)	&($I$ mag)	&	\\
 %\hline
 %& Confirmed variables\\

  \hline  
   
    V3			&5220.1629		&0.736051	&0.7350907	&0.73602	&16.05	&15.47	&0.93	&0.62	& RR0	\\
    V4			&5634.1150		&0.633806	&0.6341531	&0.63492	&16.11	&15.57	&1.01	&0.54	& RR0	\\ 
    V5			&5634.1152		&0.668918	&0.6683208	&$-$		&15.72	&15.43	&0.61	&0.47	& RR0	\\
    V6			&5989.1566		&0.347110	&0.3387880	&0.33522	&16.16 	&15.80 	&0.52	&0.42	& RR1	\\
    V7			&$-$				&$-$			&$-$			&$$	     	&$13.65^b$ &$12.90^b$ 	&$\sim 0.70$&$\sim 0.54$	& Long-term\\
    V9			&5989.1426		&0.359830	&0.3616000	&$-$		&$16.07^b$ &$15.63^b$ 	&0.36	&0.21	& RR1$^a$ \\
    V10		&5220.1321		&0.728755 	&0.7279145	&$-$		&15.03	&14.46 	&0.34	&0.25	& RR0 \\
    V11		&5249.2110		&0.823500	&0.8199846	&$-$		&$16.01^b$ &$15.31^b$ 	&0.51	&0.51	&RR0\\
    V12		&5220.3042		&0.324243	&0.3234196	&$-$		&$16.11^b$ &$-$ 	&0.64	&0.64	&RR1\\
    V13		&4840.1208		&0.689388	&0.6906617	&$-$		&16.06	&$-$ 	&0.92	&$\sim 1.50$	& RR0\\
    V14		&5220.1778		&0.323733 	&$-$			&$-$		&16.15	&15.78	&0.28	&0.26	& RR1	\\

\hline \hline
  \end{tabular}
  \caption{Data for all confirmed variable stars in NGC 1904. Periods are determined using phase dispersion minimisation (PDM) and the string length method (see text). $<V>$ and $<I>$ are intensity-weighted mean magnitudes. $^a$ potentially RRd, see text. $^b$ error-weighted mean magnitudes, rather than intensity-weighted magnitudes. Coordinates for all confirmed variables are given in \Tab{tab:coordinates}. \label{tab:knownvar}}
  \end{center}
\end{table*}
% =====================================================

% =====================================================
\begin{table}
\begin{center}
  \begin{tabular}{ccc}

     \hline
    \#		 			&RA			& DEC \\
  \hline  
    V3					&05:24:13.53	&-24:32:28.9 \\
    V4					&05:24:17.76	&-24:32:16.1 \\
    V5					&05:24:10.21	&-24:31:02.9 \\
    V6					&05:24:06.01	&-24:29:32.5 \\
    V7					&05:24:12.67	&-24:31:41.7 \\
    V9					&05:24:12.56	&-24:31:52.6 \\
    V10				&05:24:12.10	&-24:31:34.2 \\
    V11				&05:24:11.92	&-24:31:34.1 \\
    V12				&05:24:11.36	&-24:31:27.3 \\
    V13				&05:24:10.57	&-24:31:11.1 \\
    V14				& 05:24:07.76	&-24:31:00.0 \\
\hline \hline
  \end{tabular}
  \caption{J2000.0 equatorial coordinates for all confirmed variable stars in NGC 1904, derived from our astrometric fit to the $V-$band reference image. \label{tab:coordinates}}
  \end{center}
\end{table}
% =====================================================

\Tab{tab:knownvar} lists all confirmed variables in NGC 1904, including the 10 already known variables, and their coordinates obtained from the astrometric fit to our reference image are given in \Tab{tab:coordinates}. A finding chart of the cluster showing the location of the variables is shown in \Fig{fig:fchart}. When possible, the variable type is also given in the last column of the table. We performed a period search using both phase dispersion minimisation (PDM, \citealt{stellingwerf78}) and the ``string length" method \citep{lafler65}. The periods we find offer an improvement in precision compared to the periods reported by \cite{amigo11}, thanks to our long baseline. Indeed, using the periods of \cite{amigo11} with our data results in poorly phased light curves.

\begin{figure*}
  \centering
  \includegraphics[width=12cm, angle=0]{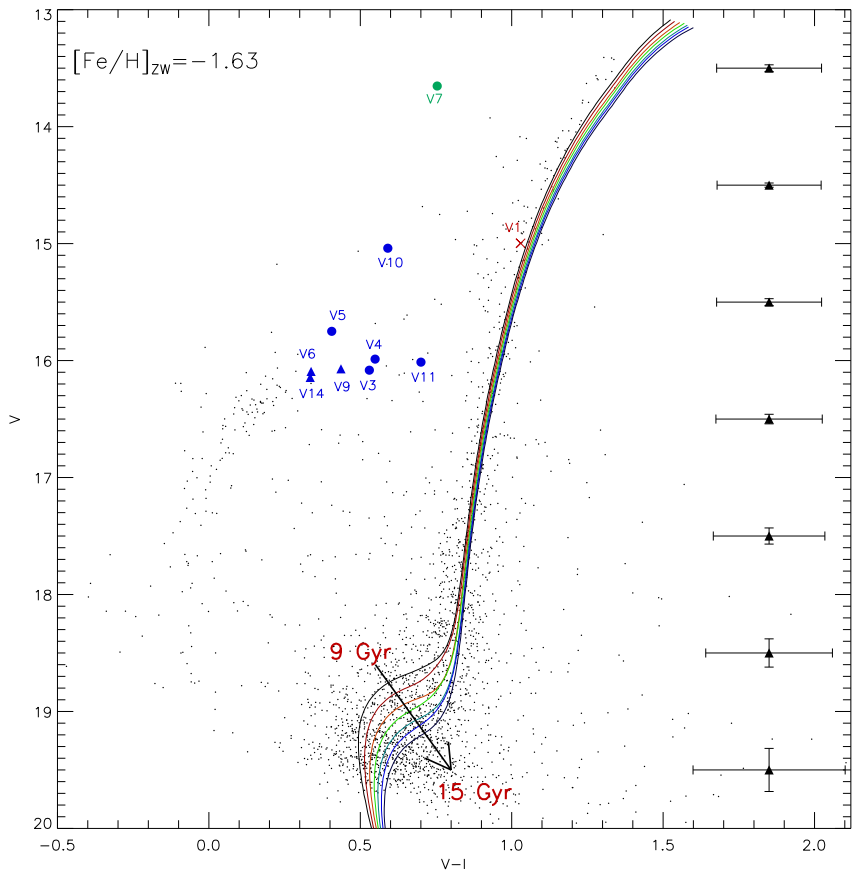}

  \caption{$V-I, V$ colour-magnitude diagram of NGC1904 extracted from the reference images. The locations of the confirmed RR Lyrae variables with $V$ and $I$ photometry are marked with blue filled circles (RR0 variables) and blue filled triangles (RR1 variables), V7 with a green filled circle, and V1, for which we do not find evidence of variability, with a red cross. Note that V5 and V10 lie above the horizontal branch because they are significantly blended, as discussed in the text. Error bars for various magnitude levels are shown near the right edge of the plot. Also shown are an interpolation of the \cite{vandenberg06} isochrones for a metallicity of $\fehzw=-1.63$ (see Sec. \ref{sec:clustermet}) at ages of 9, 10, 11, 12, 13, 14 and 15 Gyr (marked by lines of different colours; see Sec. \ref{sec:clusterage}). \label{fig:cmd}}

\end{figure*}

\begin{figure}
  \centering
  \includegraphics[width=8cm, angle=0]{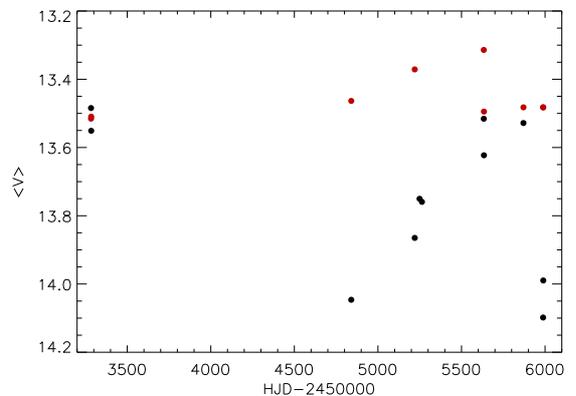}

  \caption{Average $V$-band magnitude of V7 for each observing night, showing some long-term variability of an undetermined nature. The $V$-band data is plotted in black, while the $I$-band data is plotted in red, with a fixed offset to bring the two data sets onto the same plot. The variations in $(V-I)$ therefore appear to indicate a change with time in the colour of V7, although the $I-$band data is poor. The $x$-axis is given in HJD-2450000. \label{fig:v7}}

\end{figure}

V7 shows some long-term variability and no short-term variability, and is therefore unlikely to be an RR Lyrae variable; furthermore, it was not detected as such by the work of \cite{amigo11}, and its position on our CMD (Fig. \ref{fig:cmd}) also makes it an unlikely RR Lyrae candidate. A plot of average nightly magnitude against time for V7 is shown in \Fig{fig:v7}, showing the long-term variation.

\begin{figure*}
  \centering
  \includegraphics[width=14cm, angle=0]{fig/colourcode.ps}   \\
  \includegraphics[width=6cm, angle=0]{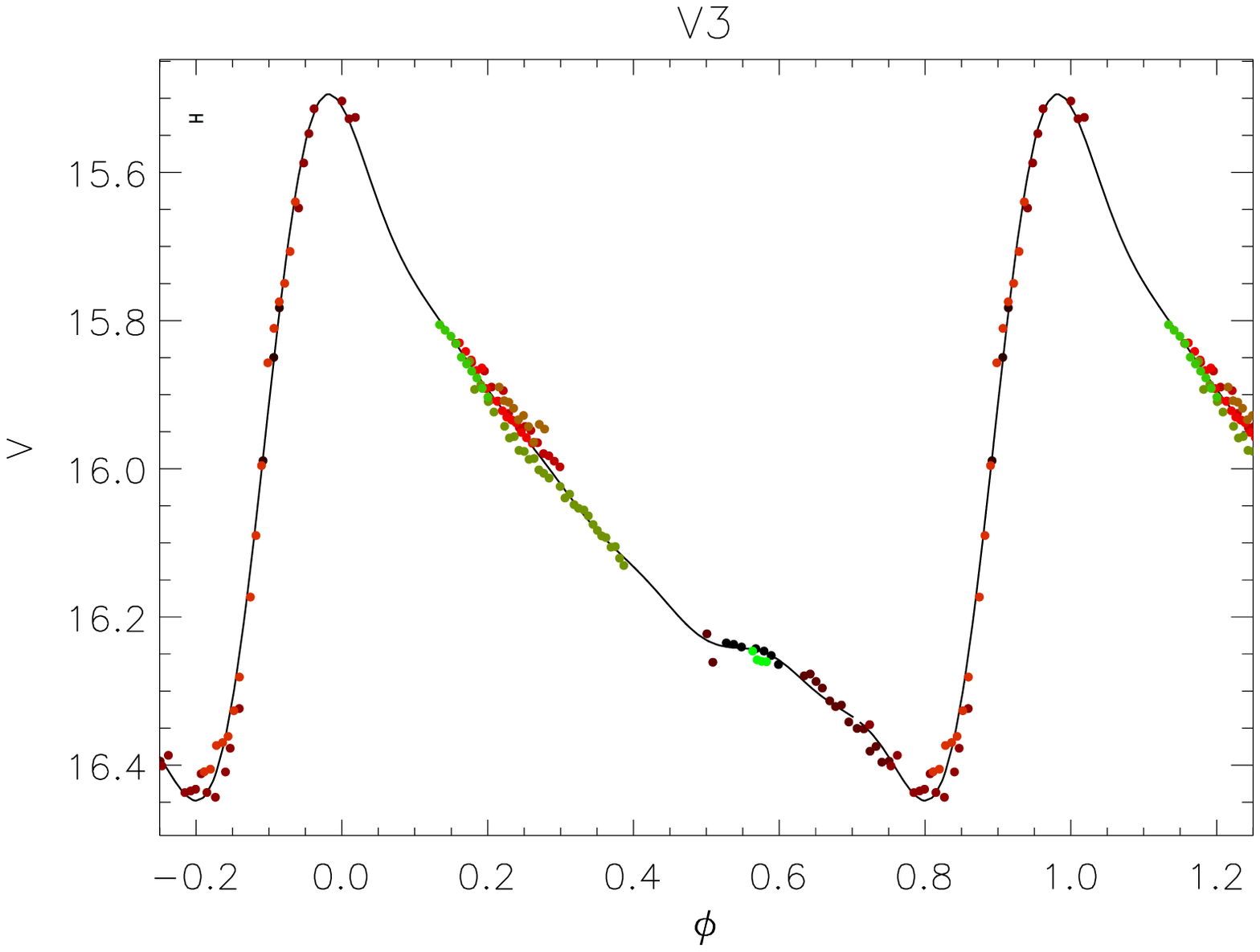}
  \includegraphics[width=6cm, angle=0]{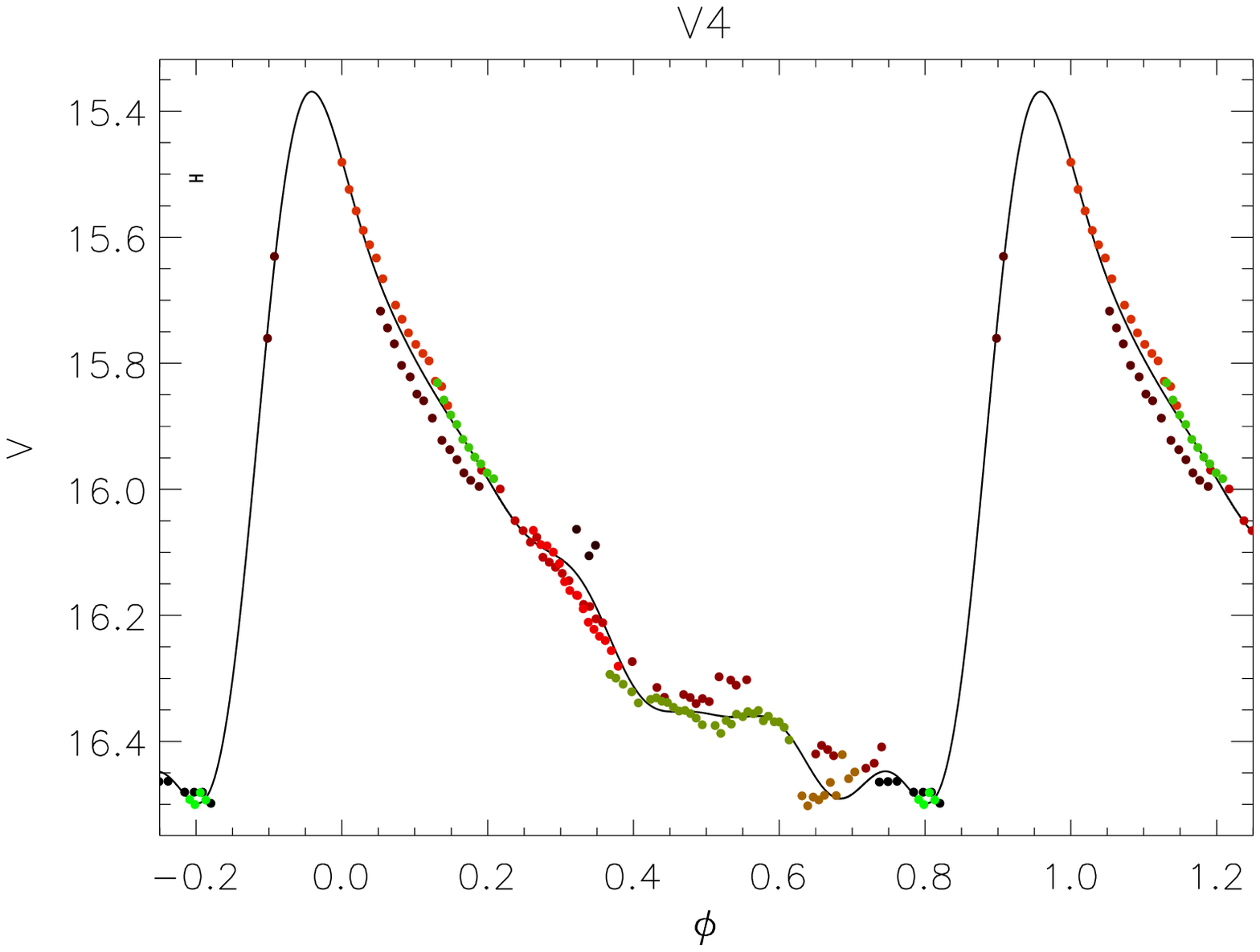}
  \includegraphics[width=6cm, angle=0]{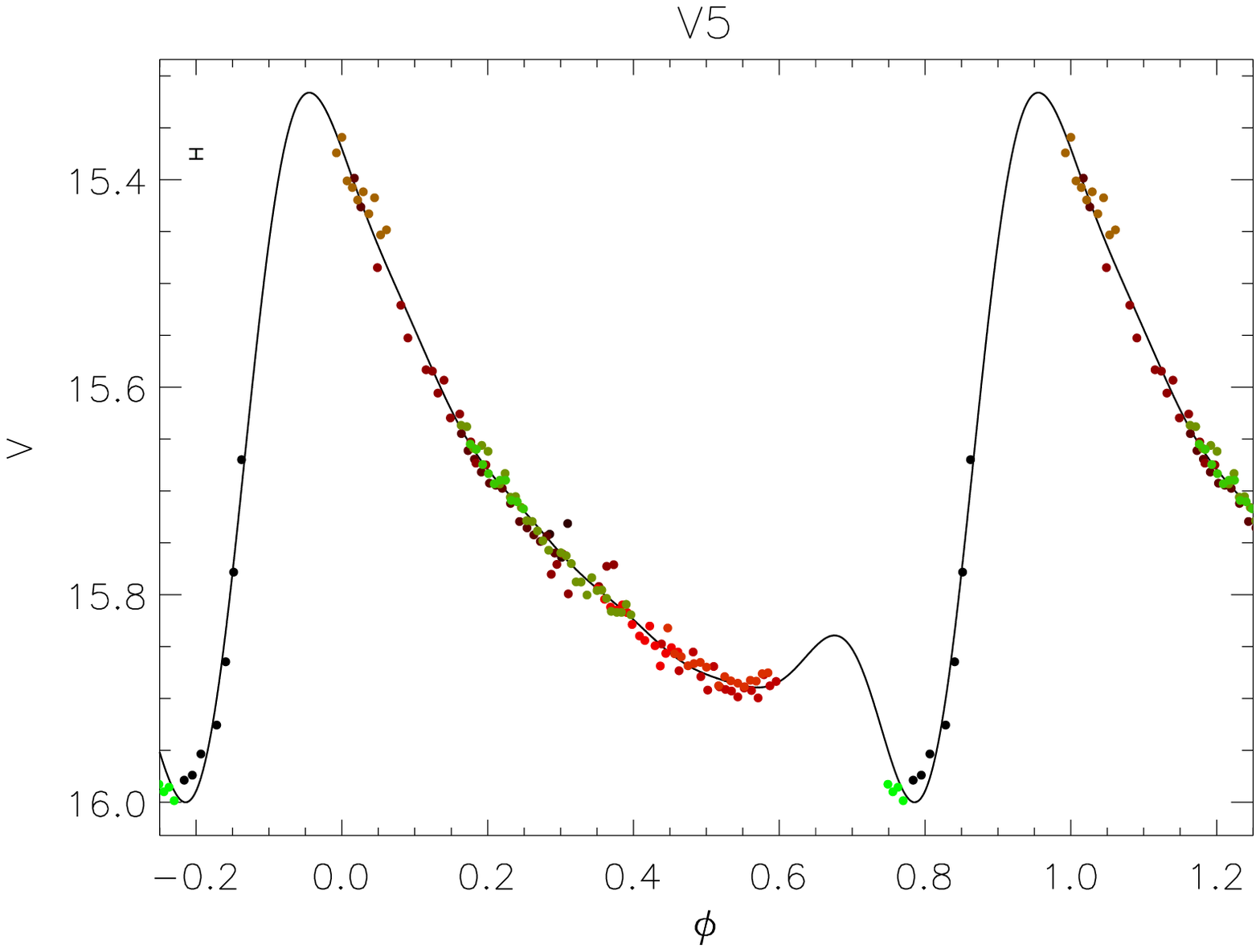}
  \includegraphics[width=6cm, angle=0]{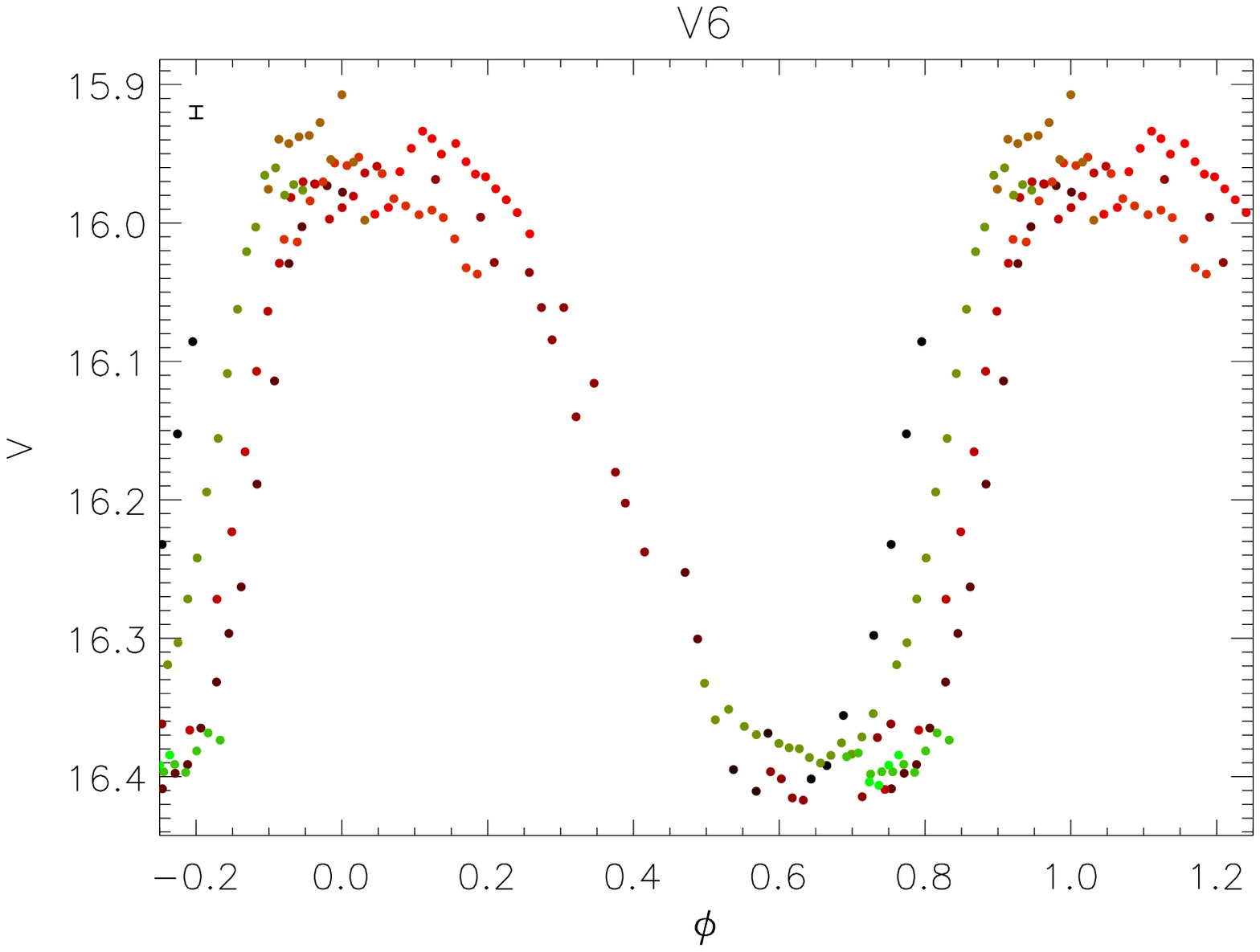}
  \includegraphics[width=6cm, angle=0]{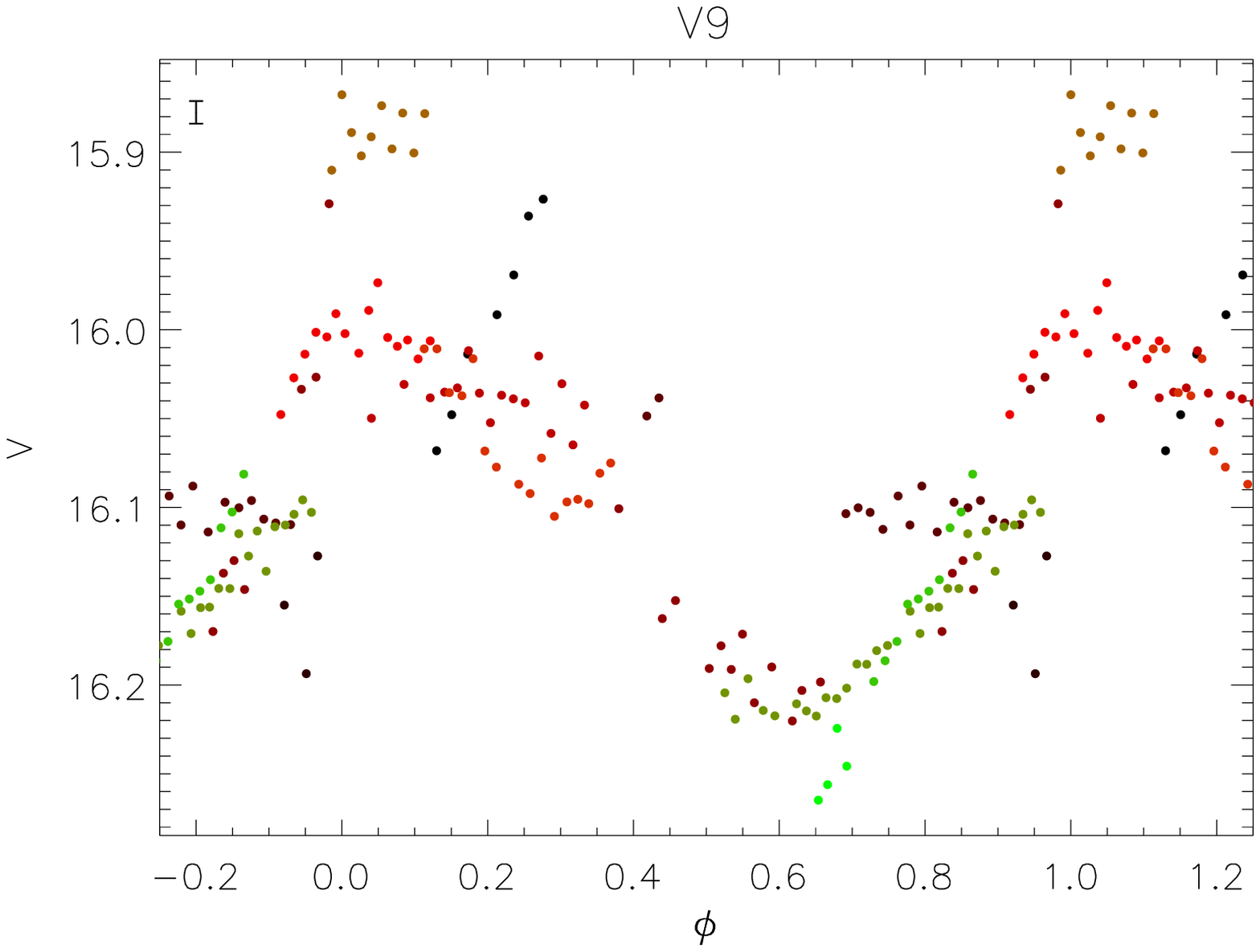}
  \includegraphics[width=6cm, angle=0]{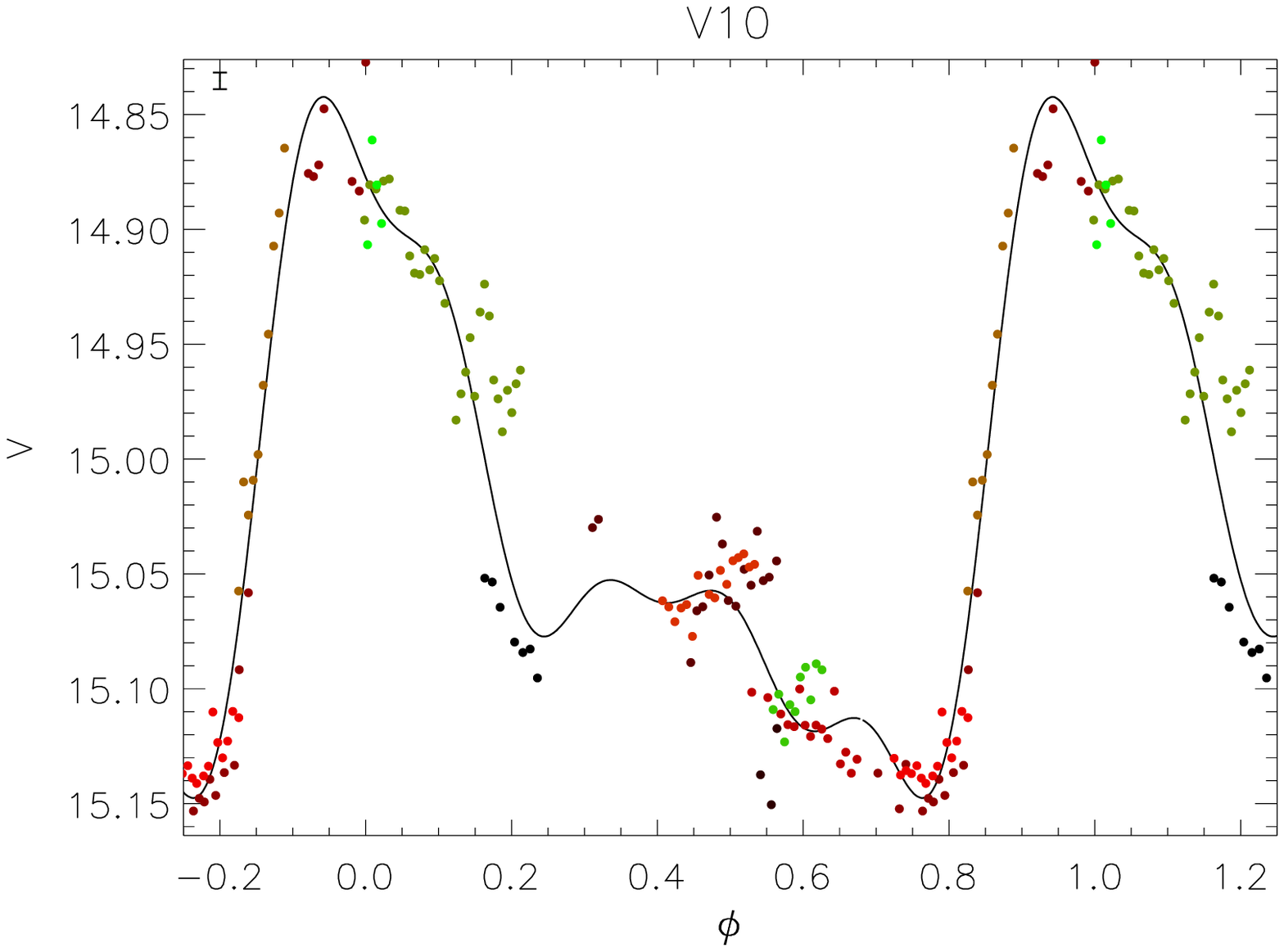}   
  \includegraphics[width=6cm, angle=0]{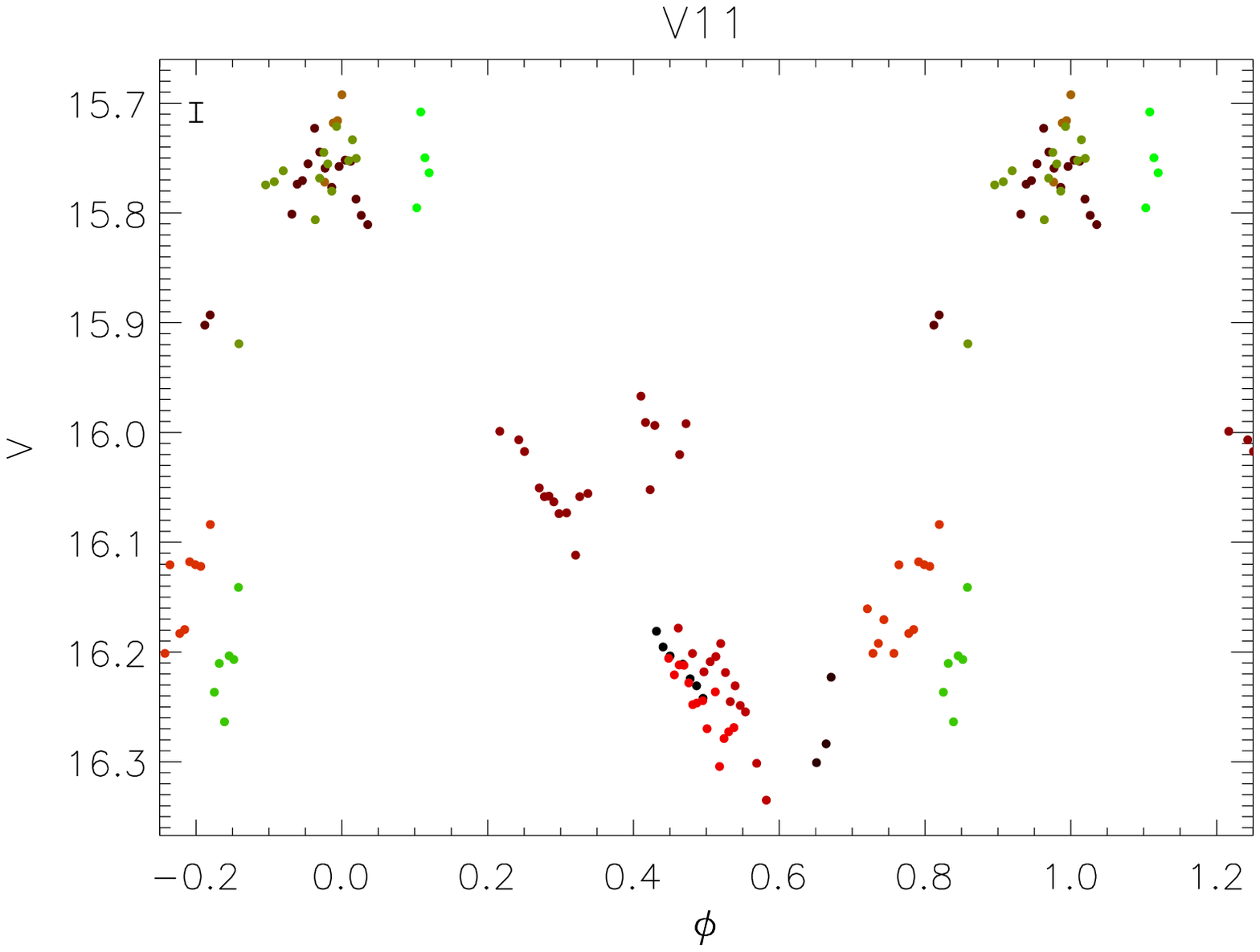}   
  \includegraphics[width=6cm, angle=0]{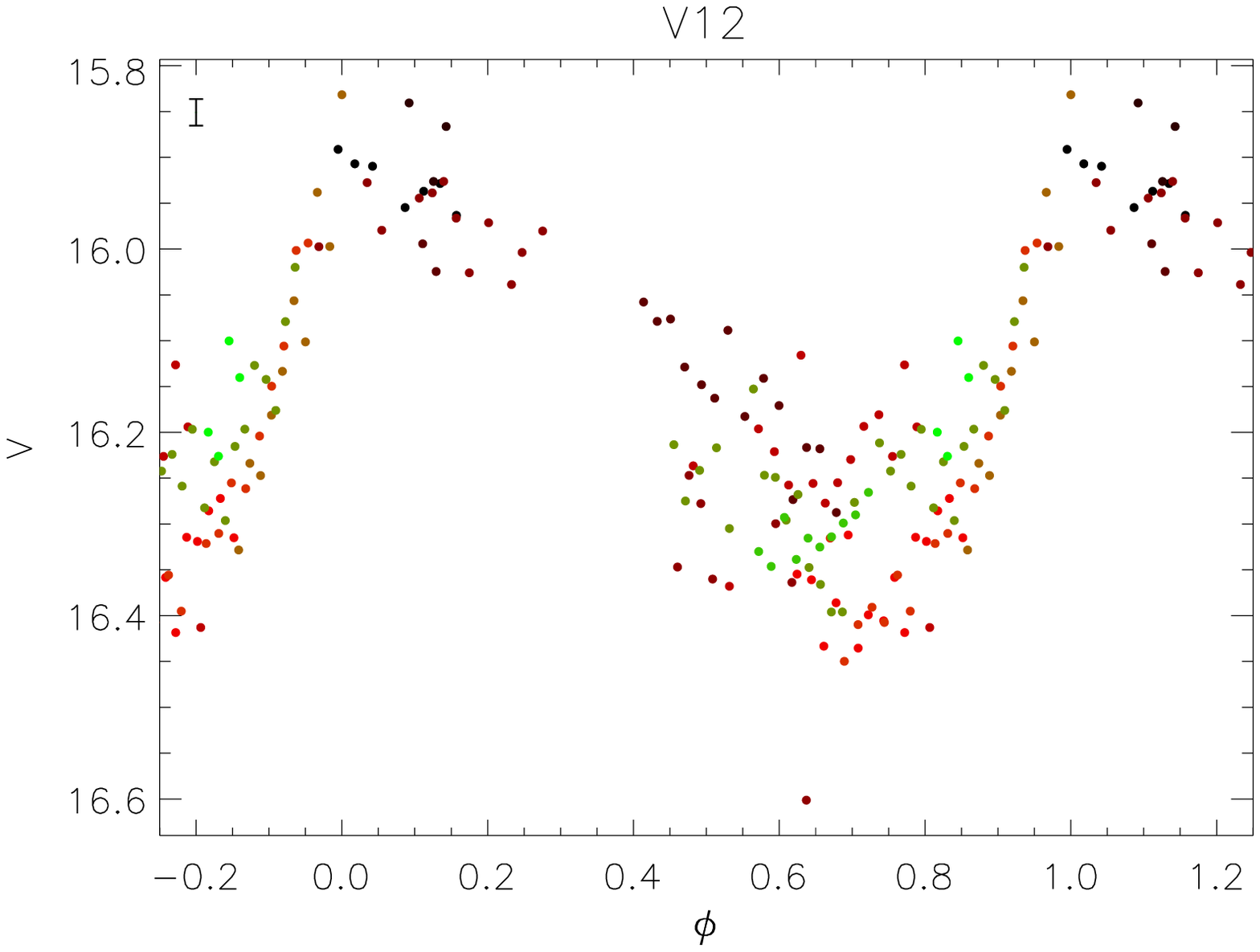}   
  \includegraphics[width=6cm, angle=0]{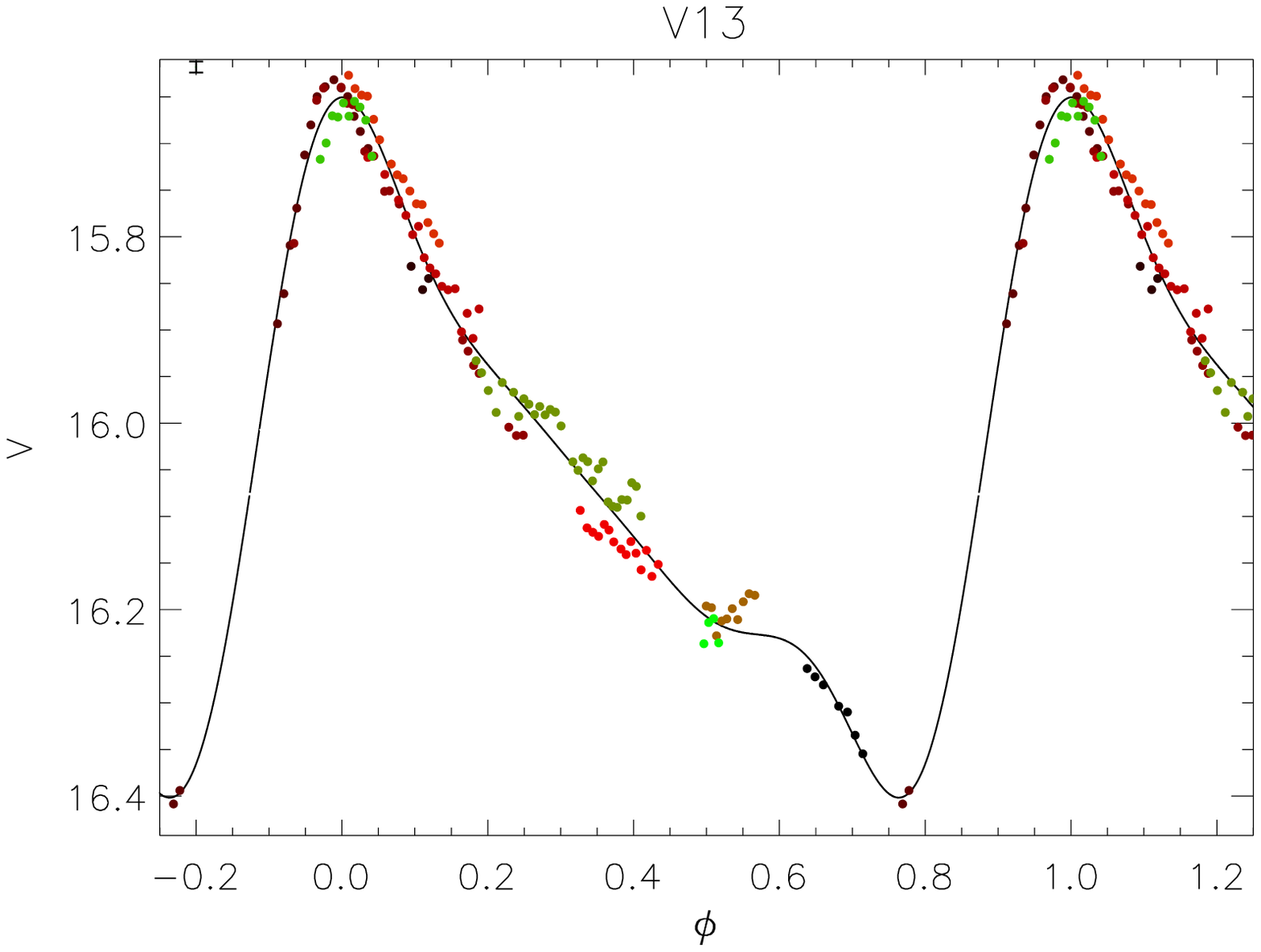}   
  \includegraphics[width=6cm, angle=0]{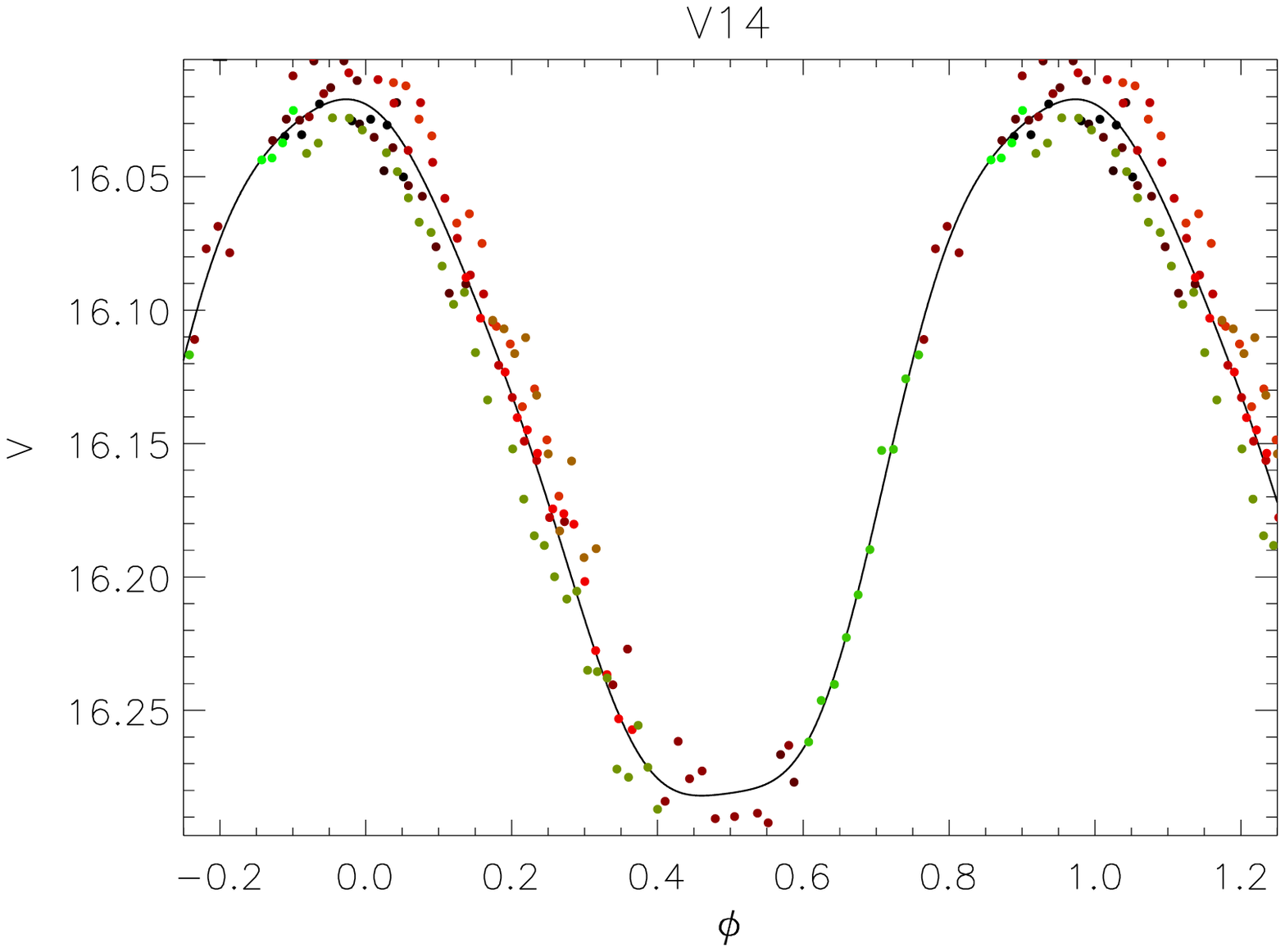}   

  \caption{Phased V-band light curves for the variables with a period estimate. Different colours are used for different nights, with the colour coding given in the form of a colour bar spanning the whole time range of the observations (top panel). Also shown is the Fourier decomposition fit to the lightcurves, for those stars for which this fit was performed (see text). The median data error bar is plotted in the top left corner of each plot. \label{fig:lc_V}}

\end{figure*}

\Fig{fig:lc_V} shows phased $V$ light curves of all confirmed variables for which a period could be estimated. $I$-band light curves were also obtained, except for V12 as the photometry was not good enough, with poor seeing and the star being located in the centre of the cluster, and V13 whose location next to a bright star and poor seeing prevented us from extracting a good quality $I$-band light curve. We do not show our $I-$band light curves here, because they are noisy and were not significant to the analysis presented in this paper. However, the light curves in both $V$ and $I$ are available in electronic format.

% =====================================================
% =====================================================

% =====================================================
% =====================================================

The phased light curves of both V4 and V6 suggest that they might exhibit the Blazhko effect \citep{blazhko1907}, leading to variations in the period and amplitude of the variation in brightness. The effect is weak in V4, but very pronounced in the case of V6, and we suggest that this is the reason why \cite{amigo11} also had issues phasing the light curve for this object. Unfortunately our time coverage does not allow us to estimate the period and magnitude of the Blazhko effect for any objects in this study, but we note the similarity of the light curves of V4 and V6 to those of many of the RR1 variables with a detected Blazhko effect discovered by \cite{arellano12}.

We see from the reference image residuals after PSF fitting that V5 and V10 are both blended in our data with stars of similar brightness, mostly due to the poor seeing of $\sim 1.5''$ in our data. This means that although the shape of the light curves is not affected (see \Sec{sec:dia}), the reference flux is overestimated for these objects; their position on the CMD is also affected, as can be seen in Fig. \ref{fig:cmd}). This is taken into account in the analysis when deriving properties of the variable stars. 

However, we note here that the position of V5 and V10 on our CMD could also be due to NGC 1904 having a prominent blue HB morphology. Due to this, some of the RR Lyrae in this cluster could be the progeny of blue HB stars during their evolution to the asymptotic giant branch (AGB) phase. Such stars cross the instability strip at higher luminosities and could therefore appear on our CMD where we find V5 and V10. The CMD obtained for this cluster by \cite{piotto02} indeed shows such stars present in NGC 1904. However, we also examined HST archival data on this cluster and verified that these two stars are indeed blended within the FWHM of our reference images.

Finally, V9 is also not well phased with our best period estimate, and from the shape of its light curve we suggest, like \cite{amigo11}, that it could be a double-period (RRd) variable.

% =====================================================
% =====================================================

\subsection{Detection of new variables}

In order to detect new variables, we employed three methods: firstly, we constructed a stacked image $S$ consisting of the sum of the absolute values of the deviations $D$ of each image from the convolved reference image, divided by the pixel uncertainty $\sigma$, so that

\begin{equation}
S_{ij} = \sum_{k}\frac{|D_{kij}|}{\sigma_{kij}} \, .
\end{equation}

\begin{figure}
  \centering
\includegraphics[width=8cm, angle=0]{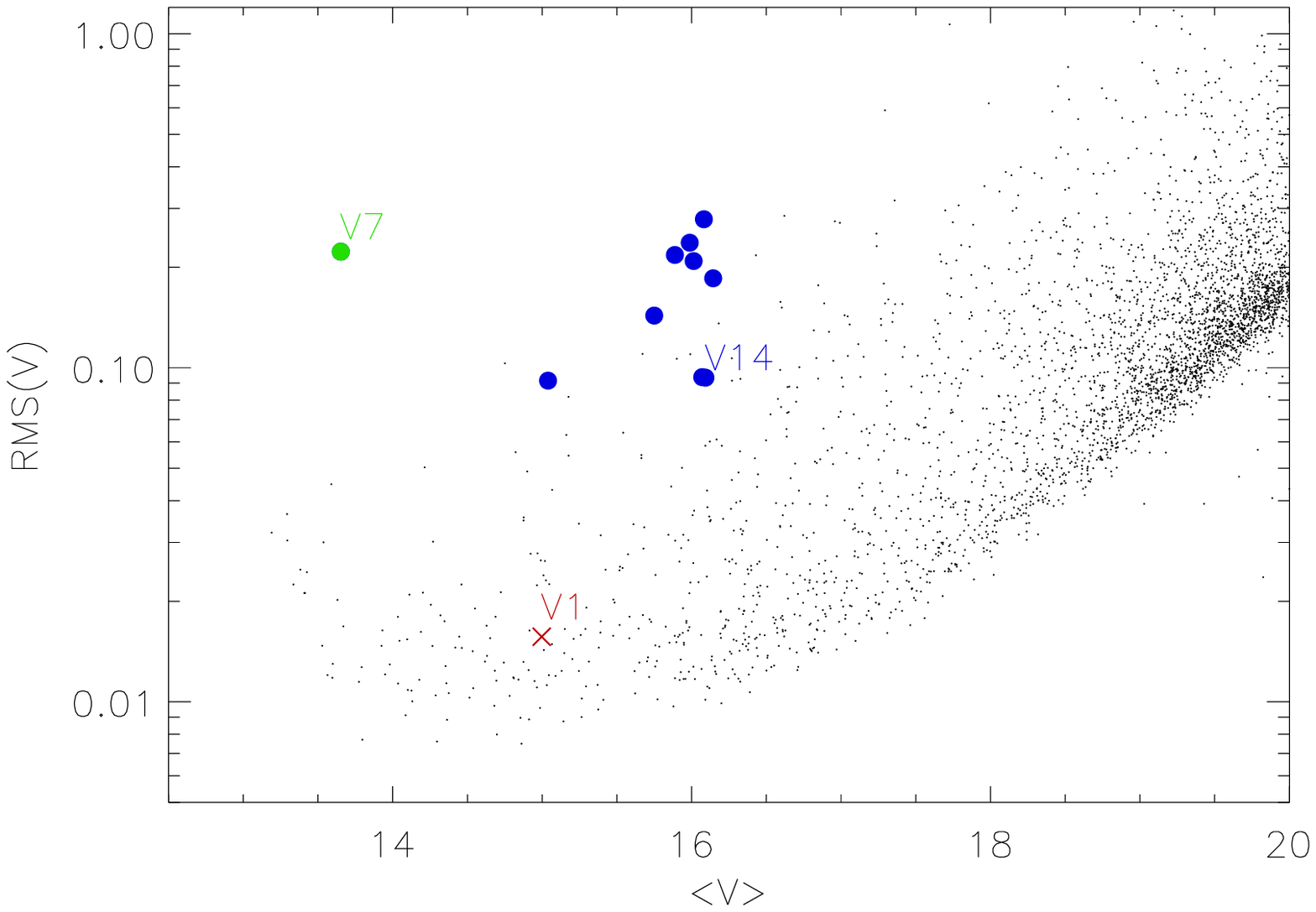}
\includegraphics[width=8cm, angle=0]{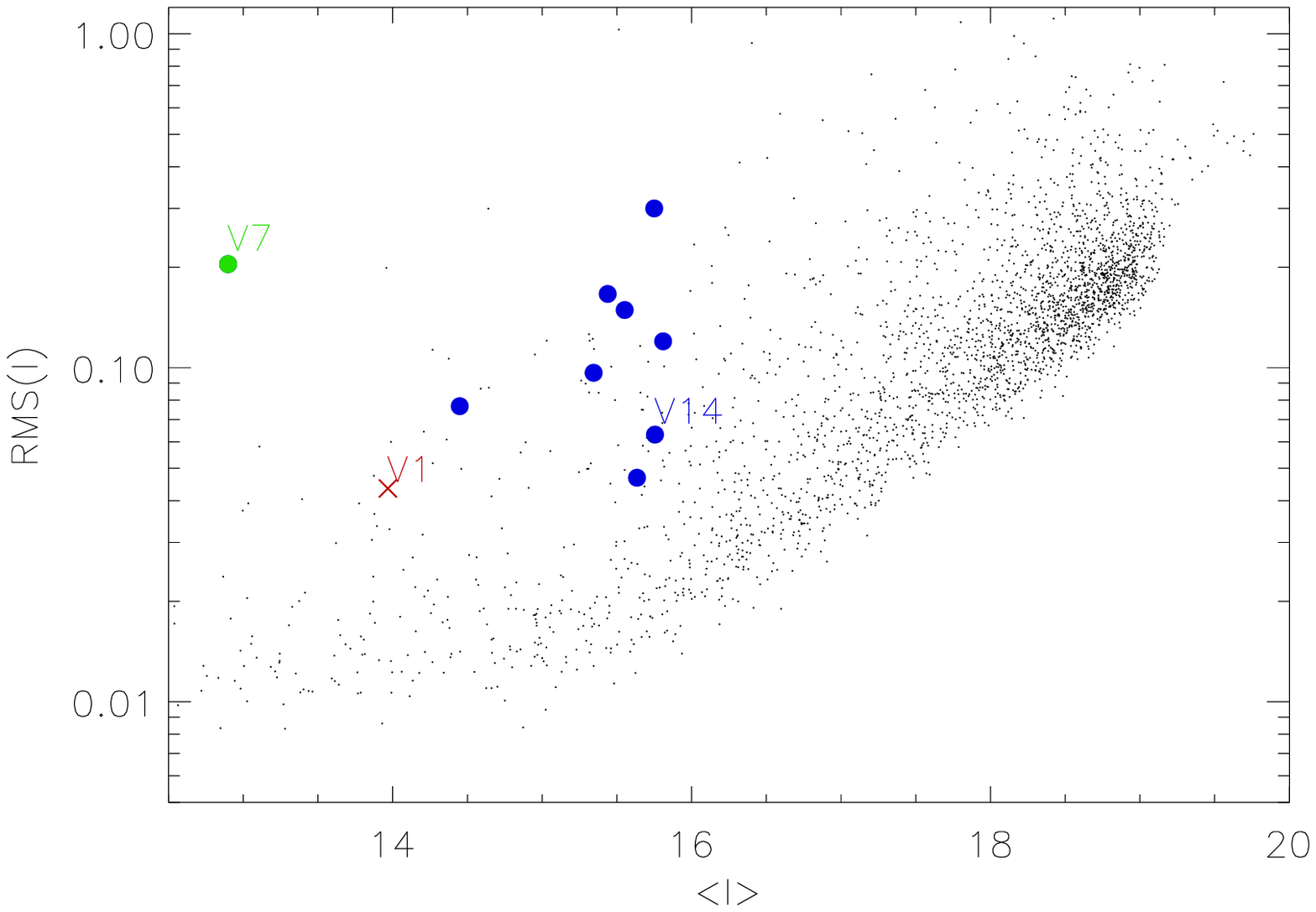}

\caption{Plot of root mean square magnitude deviation versus mean magnitude for all stars for which photometry was obtained. RR Lyrae variables are plotted as blue filled circles, V7 as a green filled circle, and V1, for which we do not find evidence of variability, is shown as a red cross. Plots are for the $V-$band (top) and $I-$band (bottom). \label{fig:rms}}

\end{figure}

Stars which consistently deviate from the reference image then stand out on this stacked image. Using this method, we discovered V14, a previously unpublished RR1 variable (see \Tab{tab:knownvar}). Secondly, we inspected the light curves of objects which stood out on a plot of root mean square magnitude deviation versus mean magnitude, shown in \Fig{fig:rms}.

\begin{figure}
  \centering
\includegraphics[width=8cm, angle=0]{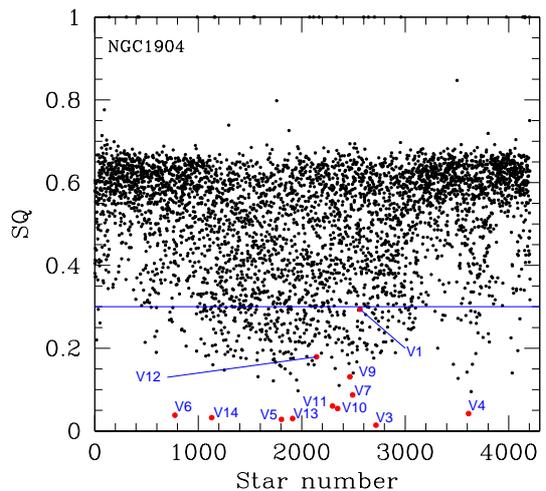}

\caption{Distribution of the $S_Q$ statistic (see text) applied to the $V$-band light curves of all stars in our sample (black filled circles). A blue horizontal line marks the threshold of $S_Q=0.3$ below which we inspected the light curves for periodic variability. The locations of the confirmed variables in our sample are marked with red filled circles; V7 is not shown as it does not appear to be periodic, but rather shows long-term variability. \label{fig:sq}}

\end{figure}

Finally, we also searched for variables by computing the string-length $S_Q$ statistic \citep[e.g.][]{dworetsky83} for all our light curves. We inspected visually all the light curves of stars with $S_Q < 0.3$, where the threshold value of 0.3 was chosen by inspecting the distribution of $S_Q$ (see \Fig{fig:sq}). However, this did not reveal any additional RR Lyrae or other variable candidates. 

We estimated the period of V14 using both PDM and the string-length method, but like the previously known variables V4 and V6, we suggest that V14 is affected by the Blazhko effect, which explains the difficulty in phasing our light curve.

% =====================================================
% =====================================================

% =====================================================
% =====================================================

%\subsection{SX Phoenicis stars}
%
%We searched for SX Phoenicis in NGC 1904 by inspecting light curves that stood out on \Fig{fig:rms} and by computing the $S_B$ index, defined as
%
%\begin{equation}\label{eq:sb}
%S_B =
%\end{equation}
%
%We inspected all light curves for stars with a value of $S_B$ smaller than ... to determine whether any of them showed obvious periodic signals. We also carried out a period search on these candidates, limiting our period search to the range $0.01 < P < 0.2$, i.e. the typical period range of SX Phoenicis stars. Using these techniques, we found no SX Phoenicis stars in NGC 1904 within the limits of our photometric precision.

\section{Variable Properties}\label{sec:fourdec}

A ($V-I, V$) colour-magnitude diagram of NGC1904 extracted from the reference images is shown on \Fig{fig:cmd}, with the location of the variables marked. Although in theory a deeper CMD could be obtained by combining many images, in practice the large variations in seeing and the number of images with significantly worse seeing than the reference image makes this impractical in this case. Variable are plotted on \Fig{fig:cmd} using their error-weighted mean magnitudes.

\subsection{Fourier decomposition}

For variables with good enough phase coverage, using Fourier decomposition of their light curves can help us derive several properties of both the variables themselves and of the cluster in which they are located. We exclude V6 from the analysis that follows because of the large dispersion in our best phased light curve (possibly due to the Blazhko effect), but retain V4 and V14, for which the effect is weak. Fourier decomposition amounts to fitting light curves with the Fourier series

\begin{equation}\label{eq:fourcos}
m(t) = A_0 + \sum_{k=1}^N A_{k} \cos \left[ \frac{2\pi k}{P}(t - E) + \phi_k \right] \, ,
\end{equation}

\noindent
where $m(t)$ is the magnitude at time $t$, $N$ is the number of harmonics used in the fit, $P$ is the period of the variable, $E$ is the epoch, and $A_k$ and $\phi_k$ are the amplitude and phase of the $k^{th}$ harmonic. Epoch-independent Fourier parameters are then defined as

\begin{eqnarray}
R_{ij} &=& A_i / A_j \\
\phi_{ij} &=& j\phi_i - i\phi_j\, .
\end{eqnarray}

We chose to fit the minimum number of harmonics that yielded a good fit, taking care not to over-fit light curve features. As a check of the dependence on $N$ of the parameters we derive for each variable, we also derived stellar parameters using Fourier parameters obtained with more harmonics, and compared them to those given in the text. We found very little variation, with any changes smaller than the error bars quoted.

We list the coefficients $A_k$ we obtain for the first four harmonics in \Tab{tab:fourdec}, as well as the Fourier parameters $\phi_{21}, \phi_{31}$ and $\phi_{41}$ for the variables for which we could obtain a Fourier decomposition. We also list the deviation parameter $D_m$, defined by \cite{jurcsik96} to assess whether fit parameters can be used to derive properties of the RR Lyrae variables. Although \cite{jurcsik96} used a criterion whereby fits should have $D_m < 3$ for their empirical relations to yield reliable estimates of stellar properties, a less stringent criterion of $D_m < 5$ has been used by other authors \citep[e.g.][]{cacciari05}. Here we also adopt $D_m < 5$ as a selection criterion to estimate stellar properties, which excludes one variable, V4, from our calculations of the cluster's parameters (Sec. \ref{sec:clusterprop}).

% =====================================================
\begin{table*}
\begin{center}
  \begin{tabular}{ccccccccccc}

     \hline
    \#		&$A_0$	&$A_1$	&$A_2$ 	&$A_3$	&$A_4$	&$\phi_{21}$	&$\phi_{31}$	&$\phi_{41}$	&$N$	&$D_m$	 \\
 \hline
 RR0 \\
 \hline
V3& 16.049(1)&  0.330(1)&  0.165(1)&  0.107(1)&  0.057(1)&  4.222(9)&  8.680(14)&  6.805(19) & 6 &2.2	\\
V4& 16.111(1)&  0.408(1)&  0.195(1)&  0.134(1)&  0.062(1)&  4.031(5)&  8.318(8)&  6.599(13) & 7 &7.0	\\
V5& 15.724(1)&  0.224(2)&  0.114(2)&  0.085(1)&  0.052(1)&  4.115(23)&  8.535(28)&  6.466(32) & 6 &3.5	\\
V10& 15.027(1)&  0.102(1)&  0.073(1)&  0.018(1)&  0.019(1)&  4.045(16)&  8.644(35)&  6.543(37) & 5 &3.7	\\
V13& 16.061(1)&  0.273(1)&  0.132(1)&  0.075(1)&  0.020(1)&  4.317(15)&  8.825(22)&  7.290(49) & 4 &2.3	\\
\hline
RR1\\
\hline
V14& 16.154(1)&  0.137(1)&  0.007(1)&  0.007(1)&  0.005(1)&  4.841(140)&  3.079(121)&  2.226(152) & 4 &$-$ \\
	\hline \hline
  \end{tabular}

  \caption{Coefficients $A_k$ and selected Fourier parameters for the Fourier decomposition fit of the RR Lyrae variables in NGC1904. The number of fitted harmonics $N$ is also given. Numbers in parentheses are the 1-$\sigma$ uncertainties on the last decimal place. Note that $A_0$ for V5 and V10 is too bright and $A_1, A_2, A_3$ and $A_4$ are underestimated because these two objects are significantly blended, so we exclude them from calculations in which this parameter is used to estimate the distance to the cluster. We also list the deviation parameter $D_m$ defined by \cite{jurcsik96} to assess whether fit parameters can be used to derive properties of the RR Lyrae variables. \label{tab:fourdec}}
  \end{center}
\end{table*}
% =====================================================

\subsection{Metallicity}\label{sec:rrmet}

We use the empirical relations of \cite{jurcsik96} to derive the metallicity [Fe/H] for each of the variables for which we could obtain a successful Fourier decomposition. The relation is derived from the spectroscopic metallicity measurement of field RR0 variables, and it relates [Fe/H] to the period $P$ and the Fourier parameter $\phi^s_{31}$, where $s$ denotes a parameter obtained by fitting a \textit{sine} series rather than the cosine series we fit with \Eq{eq:fourcos}. [Fe/H] is then expressed as

\begin{equation}\label{eq:metrr0}
\mathrm{[Fe/H]_J} = -5.038 - 5.394\, P + 1.345\, \phi^s_{31}\,
\end{equation}

\noindent
where the subscript J denotes a non-calibrated metallicity, the period P is in days, and $\phi^s_{ij}$ can be calculated via 

\begin{equation}\label{eq:phis}
\phi^s_{ij} = \phi_{ij} - (i-j)\, \frac{\pi}{2}\, .
\end{equation}

We transform these to the metallicity scale of \cite{zinn84} (hereafter ZW) using the relation from \cite{jurcsik95}:

\begin{equation}\label{eq:zw}
\fehzw = \frac{\rm [Fe/H]_J - 0.88}{1.431}\, .
\end{equation}

For our only RR1 variable with a good Fourier decomposition, V14, we calculated the metallicity using the empirical relation of \cite{morgan07}, linking [Fe/H], $P$ and $\phi_{31}$:

\begin{eqnarray}\label{eq:metrr1}
\fehzw &=& 2.424 - 30.075\, P +  52.466\, P^2 \\ \nonumber
&&+ 0.982\, \phi_{31} + 0.131 \phi_{31}^2 - 4.198\, \phi_{31}P \,
\end{eqnarray}

Metallicity values calculated using \Eq{eq:metrr0} \& (\ref{eq:zw}) and (\ref{eq:metrr1}) are given in \Tab{tab:starpar}.

% =====================================================
\begin{table}
\begin{center}
  \begin{tabular}{ccccc}

     \hline
    \#		&$\fehzw$	&$M_V$	&$\log(L/L_{\bigodot})$ 	&$T_{\rm eff}$	\\
 \hline
RR0 \\
 \hline

V3&  -1.70(2) &  0.366(3) &  1.288(2) & 6232(7) \\
V4&  -1.66(2) &  0.420(2) &  1.291(1) & 6354(4) \\
V5&  -1.59(4) &  0.548(4) &  1.228(2) & 6292(12) \\
V10&  -1.71(4) &  0.564(2) &  1.195(1) & 6161(14) \\
V13&  -1.39(3) &  0.458(2) &  1.262(2) & 6287(11) \\

\hline
RR1 \\
\hline

V14&  -1.73(6) &  0.584(8) &  1.409(4) & 7293(12) \\

	\hline \hline
  \end{tabular}

  \caption{Physical parameters for the RR Lyrae variables calculated using the Fourier decomposition parameters and the relations given in the text. Numbers in parentheses are the 1-$\sigma$ uncertainties on the last decimal place. Note that $M_V$ for V5 and V10 is overestimated because these two objects are significantly blended, so we exclude them from the calculation for the distance to the cluster. \label{tab:starpar}}
  \end{center}
\end{table}
% =====================================================

\subsection{Effective Temperature}

Another property of the RR Lyrae that can be estimated through the Fourier parameters is the effective temperature. \cite{jurcsik98} derived empirical relations linking the $(V-K)_0$ colour to $P$ as well as several of the Fourier coefficients and parameters:

\begin{eqnarray}
(V-K)_0 &=&  1.585 + 1.257\, P - 0.273\, A_1 - 0.234\, \phi^s_{31} \\ \nonumber
&&+ 0.062\, \phi^s_{41} \\\label{eq:teffrr0}
\mathrm{log}\,T_{\rm eff} &=& 3.9291 - 0.1112\,(V-K)_0 \\ \nonumber
&& - 0.0032 \, \mathrm{[Fe/H]_J} \, .
\end{eqnarray}

For RR1 variables, we use a corresponding relation derived from theoretical models by \cite{simon93} to calculate $\teff$,

\begin{equation}\label{eq:teffrr1}
\log\,\teff = 3.7746 - 0.1452 \log\, P + 0.0056\, \phi_{31} \, .
\end{equation}

The temperatures we derive using these relations are given in \Tab{tab:starpar}; note that because V5 and V10 are significantly blended, the temperatures we calculated for them are underestimated. As noted by \cite{bramich11}, there are several caveats to deriving temperatures with \Eq{eq:teffrr0} and (\ref{eq:teffrr1}). The values of $\teff$ derived for RR0 and RR1 variables are on different absolute scales. Furthermore, the effective temperatures we derive here deviate systematically from the relations predicted by evolutionary models of \cite{castelli99} or the temperature scales of \cite{sekiguchi00}; however this was already noted in previous work \citep{arellano08a, arellano10, bramich11}, and the temperatures given here are a useful comparison with previous analogous studies on other clusters.

\subsection{Absolute Magnitude}\label{sec:rrmag}

We use the empirical relations of \cite{kovacs01} to derive V-band absolute magnitudes for the RR0 variables. The relation links the magnitude to Fourier coefficients through

\begin{equation}\label{eq:mvrr0}
M_V = -1.876 \log P - 1.158\, A_1 + 0.821 \, A_3 + K_0 \, ,
\end{equation}

\noindent
where $K_0$ is a constant. As in \cite{bramich11} and \cite{arellano10}, we choose a value of $K_0=0.41$ mag to be consistent with a true LMC distance modulus of $\mu_0=18.5$ mag \citep{freedman01}. For RR1 variables, we use the relation of \cite{kovacs98},

\begin{equation}\label{eq:mvrr1}
M_V = -0.961\, P - 0.044\, \phi^s_{21} - 4.447 \, A_4 + K_1 \, ,
\end{equation}

\noindent
where $K_1$ is a constant, for which we choose a value of 1.061 with the same justification as for our choice of $K_0$.

We also converted the magnitudes to luminosities using

\begin{equation}\label{eq:logl}
\log\left( L/L_{\bigodot} \right) = -0.4\, \left[M_V + B_C(\teff) - M_{\rm bol, \bigodot }\right] \, ,
\end{equation}

\noindent
where we use $M_{\rm bol, \bigodot }=4.75$, and $B_C(\teff)$ is a bolometric correction which we determine by interpolating from the values of \cite{montegriffo98} and using our best-fit value of $\teff$. We report values of $M_V$ and $\log(L/L_{\bigodot})$ for the RR0 and RR1 variables in \Tab{tab:starpar}, and also note that, because V5 and V10 are heavily blended, their derived value of $M_V$ is likely to be overestimated; we therefore do not use them to estimate the distance to the cluster.

\section{Cluster Properties}\label{sec:clusterprop}

\subsection{Oosterhoff Type}

Although the number of variables for which the period could be determined is small, we calculate the mean periods of the RR0 and RR1 variables as $<P_{\rm RR0}>=0.71 \pm 0.07$ and $<P_{RR1}>=0.33 \pm 0.02$. RR1 variables account for 40\% of all RR Lyrae stars in this cluster.

\begin{figure}
  \centering
  \includegraphics[width=8cm, angle=0]{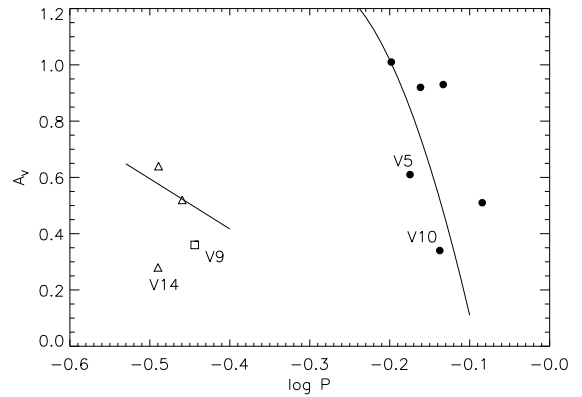}

  \caption{Bailey diagram of the $V$-band light curve amplitude versus the logarithm of the period for RR Lyrae stars in NGC 1904. RR0 variables are plotted as filled circles, and RR1 variables as open triangles; V9 is plotted as an open square as it is a potential double-mode RR Lyrae star. We also mark the locations of V5 and V10, whose amplitude is underestimated because of blending. Also plotted as solid lines are the relations of \cite{cacciari05} for evolved stars; these are obtained by applying a shift of +0.06 to $\log\, P$ in the relations derived for Oosterhoff type I cluster M3. \label{fig:bailey}}

\end{figure}

The value of $<P_{\rm RR0}>$ and the high fraction of RR1 variables makes this cluster an Oosterhoff type II, in agreement with previous classifications \citep[e.g.][]{amigo11}. We also show the distribution of variables in a Bailey diagram on \Fig{fig:bailey}, and confirm that the distribution is consistent with an Oosterhoff type II cluster; as found by \cite{cacciari05}, type II clusters have $\log\, P - A_V$ tracks that correspond to the tracks of evolved stars in the ``prototype" type I cluster, M3. Although the metallicity value we derive for NGC 1904 makes it one of the most metal-rich Oosterhoff type II clusters \citep[e.g.][]{sandage93}, there are several other known type II Galactic clusters with similar metallicities such as M2 \citep{lee99} and M9 \citep{clement99}. Furthermore, \cite{lee99} and \cite{clement99b} both concluded that the Oosterhoff dichotomy was due to evolution rather than metal content; we therefore do not conclude that our metallicity value and Oosterhoff classification are contradictory.

% =====================================================
% =====================================================

\cite{vanalbada73} suggested that in Oosterhoff type II clusters, RRc evolve to lower temperatures, but that there is a hysteresis effect so that mode switching occurs at lower temperatures for stars evolving to lower temperatures than for stars evolving to higher temperatures. Indeed for NGC 1904, we find a mean temperature for RR0 stars (excluding V4, V5 and V10) of $6260 \pm 39$K, which is lower than the means found for Oosterhoff type I clusters, e.g. $6418 \pm 10$K for NGC 6981 \citep{bramich11}, $6633 \pm 257$K for NGC 4147 \citep{arellano04}, $6465 \pm 22$K for M 5 \citep{kaluzny00}, $6494 \pm 35$K for NGC 1851 \citep{walker98}, $6619 \pm 64$K for NGC 6171 \citep{clement97}.

\subsection{Metallicity}\label{sec:clustermet}

Although \cite{sandage06} found that there is a correlation between the mean period of the RR0 variables and the cluster metallicity for Oosterhoff type I clusters, this relation breaks down when $<P_{\rm RR0}>$ is larger than $\sim 0.6$, i.e. for Oosterhoff type II clusters \citep{clement01}. 

On the other hand, we can use the metallicities derived in \Sec{sec:rrmet} to estimate the cluster's metallicity. Taking an average of the RR Lyrae metallicities listed in \Tab{tab:starpar}, and assuming as in \cite{bramich11} that there is no systematic offset between metallicity estimates for the different types of variables, we find a mean metallicity of $\fehzw=-1.63 \pm 0.14$\footnote{Note that the values in \Tab{tab:starpar} are rounded off, which explains why the mean of those values is not equal to the value we derive for the cluster metallicity.}. This is in good agreement with values in the literature, listed in \Tab{tab:lit_met}. Note that V4 was excluded from this calculation as it has $D_m > 5$, meaning that its metallicity estimate might be unreliable. The error bars given here and for all cluster properties are the scatter around the mean value. 

We also transform this metallicity value to the scale of \cite{carretta09}, who derived a new metallicity scale based on GIRAFFE and UVES spectra of red giant branch (RGB) stars in 19 globular clusters. The transformation from the ZW to the UVES \citep{carretta09} scale is given as

\begin{equation}\label{eq:fehuves}
\fehuves = -0.413 + 0.130\,\fehzw - 0.356\,\fehzw^2\, .
\end{equation}

Using this we find a metallicity for NGC 1904 of $\fehuves=-1.57 \pm 0.18$, in excellent agreement with the value found for this cluster by \cite{carretta09} of $\fehuves = -1.58 \pm 0.03$.

% =====================================================
\begin{table*}
\begin{center}
  \begin{tabular}{ccc}

     \hline
    Reference		&$\fehzw$	&Method	\\
 \hline
This work			&-1.63 $\pm$ 0.14	& Fourier decomposition of RR Lyrae light curves\\
\cite{kraft03}		&-1.68		& EW of Fe II spectral lines \\
%\cite{rosenberg99}	&-1.37 $\pm$ 0.05	&	\\
\cite{carretta97}	&-1.36 $\pm$ 0.09	& Spectroscopy of red giants	\\
\cite{kravtsov97}	&-1.76 $\pm$ 0.20	& CMD analysis\\
\cite{francois91}	&-1.46 $\pm$ 0.15	&  $Q_{39}$ spectral index\\
\cite{gratton89}		&-1.42 $\pm$ 0.23	& Absorption line strength indices\\
\cite{brodie86}		&-1.70 $\pm$ 0.23	& Absorption line strength indices \\
\cite{smith84}		&-1.43 $\pm$ 0.20	& Corrected $Q_{39}$ index \\
\cite{nelles84}		&-1.78 $\pm$ 0.25	& Calibration of VBLUW indices \\
\cite{zinn84}		&-1.69 $\pm$ 0.09	& $Q_{39}$ spectral index \\
%\cite{bica83}		&-1.52	& C(42-45) vs [Fe/H] \\ %just recalibration
%\cite{frogel83}		&-1.65	&IR colour  \\ %just a recalibration?
\cite{zinn80}		&-1.76 $\pm$ 0.05	& $Q_{39}$ spectral index \\

	\hline \hline
  \end{tabular}

  \caption{Different metallicity estimates for NGC 1904 in the literature. \label{tab:lit_met}}
  \end{center}
\end{table*}
% =====================================================

\subsection{Distance}\label{sec:clusterage}

Having derived the absolute magnitudes of the RR Lyrae stars in \Sec{sec:rrmag}, we can use the $A_0$ parameter, which corresponds to their mean apparent V magnitude, to derive a distance modulus to NGC 1904. Note that $M_V$ for V5 and V10 is overestimated because these two objects are significantly blended, so we exclude them from this calculation, as we do with V4 because of its value of $D_m$ above our selection threshold. The mean value of the apparent $V$ band magnitudes for the RR0 variables is 16.055 $\pm$ 0.008 mag, while the mean of the absolute magnitudes is 0.412 $\pm$ 0.065 mag. This yields a distance modulus of $\mu=15.643 \pm 0.066$. Using the parameters for our lone RR1 variable (V14, see Tables \ref{tab:fourdec} \& \ref{tab:starpar}), we find $\mu=15.570$.

\cite{stetson77} used various methods to estimate the reddening towards this cluster and obtained values of $E(B-V)$ lower than 0.03 mag. \cite{ferraro92} found a reddening for NGC 1904 of $E(B-V)=0.01 \pm 0.01$ mag, while \cite{kravtsov97} estimate the reddening for this cluster to be $E(B-V)=0.035$ to 0.05 mag, depending on the metallicity of their best-fit isochrone.

Adopting a value of  $E(B-V)=0.01 \pm 0.01$ mag and a value for $R_V=3.1$ for our Galaxy, we find a mean true distance moduli of $\mu_0 = 15.612 \pm 0.066$ mag and $\mu_0 = 15.539$ mag for RR0 and RR1 variables in our samples. This yields mean distances of $13.26 \pm 0.41$ kpc and 12.87 kpc. These values are sensitive to the adopted value of $E(B-V)$, and using the largest value found in the literature, $E(B-V)=0.05$ mag yields distances of $12.52 \pm 0.38$ and $12.11$ kpc, both consistent with estimates from the literature using other methods (\Tab{tab:lit_dist}).

We checked that our distance modulus is consistent with the isochrones of \cite{vandenberg06} for the metallicity of $\fehzw = -1.63$ that we calculated in our analysis; these are overplotted on \Fig{fig:cmd} and show that our results are broadly consistent with the theoretical isochrones of that metallicity. Although isochrone fitting to the CMD could also be used to estimate the age of NGC 1904, the quality of our CMD does not allow us to do this in this case, and we can only show that the CMD is consistent with the wide range of ages reported in the literature for this cluster, from 10 to 18 Gyr (see Table \ref{tab:lit_age}).

% =====================================================
\begin{table*}
\begin{center}
  \begin{tabular}{cccc}

\hline
    Reference		&$\mu_0$ [mag]		&Distance [kpc]		&Method	\\
\hline

This work			&15.63 $\pm$ 0.06	&13.35 $\pm$ 0.35	&Fourier decomposition of RR0 light curves \\
This work			&15.54			&12.87		&Fourier decomposition of RR1 light curves \\
\cite{zoccali00}		&15.52			&12.53		&Theoretical luminosity function \\
\cite{ferraro99}		&15.63			&13.37  		&Magnitude of the horizontal branch \\
\cite{kravtsov97}	&15.57$^a$		&13.00		&CMD analaysis \\
\cite{harris96}		&$-$				&12.9		&Globular cluster catalogue \\
\cite{zinn80}		&15.62$^a$		&13.29		&$Q_{39}$ spectral index \\
\cite{rosino52}		&16.25			&17.78		&Median magnitude of RR Lyrae \\

\hline \hline
\end{tabular}

  \caption{Modulus and distance estimates for NGC 1904 in the literature. $^a$Assuming the values $E(B-V)=0.01$ and $R_V = 3.1$. \label{tab:lit_dist}}
  \end{center}
\end{table*}
% =====================================================

% =====================================================
\begin{table*}
\begin{center}
  \begin{tabular}{cccc}

     \hline
    Reference	&Age [Gyr]		&Method	\\
 \hline
\cite{koleva08}		&11.99 $\pm$ 0.57		& Spectrum fitting 	\\
\cite{koleva08}		&14.02 $\pm$ 0.25		& Spectrum fitting with hot stars \\
\cite{deangeli05}	& 10.19 $\pm$ 0.08		& Age- ZW metallicity fitting \\
\cite{deangeli05}	& 9.92 $\pm$ 0.08		& Age- CG metallicity fitting \\
\cite{santos04}		&13.2 $\pm$ 1.1		& Age-metallicity fitting  \\
\cite{zoccali00}		&14					& RGB luminosity function\\
\cite{kravtsov97}	&$16 \pm 1$			& CMD isochrone fitting \\
\cite{kravtsov97}	&$18 \pm 1$			& CMD  isochrone fitting \\
\cite{alcaino94}		&$16$				& CMD  isochrone fitting \\

	\hline \hline
  \end{tabular}

  \caption{Age estimates for NGC 1904 in the literature. \label{tab:lit_age}}
  \end{center}
\end{table*}
% =====================================================

\section{The $M_V -$ [Fe/H] relation}\label{sec:mvfeh}

\begin{figure}
  \centering
  \includegraphics[width=9cm, angle=0]{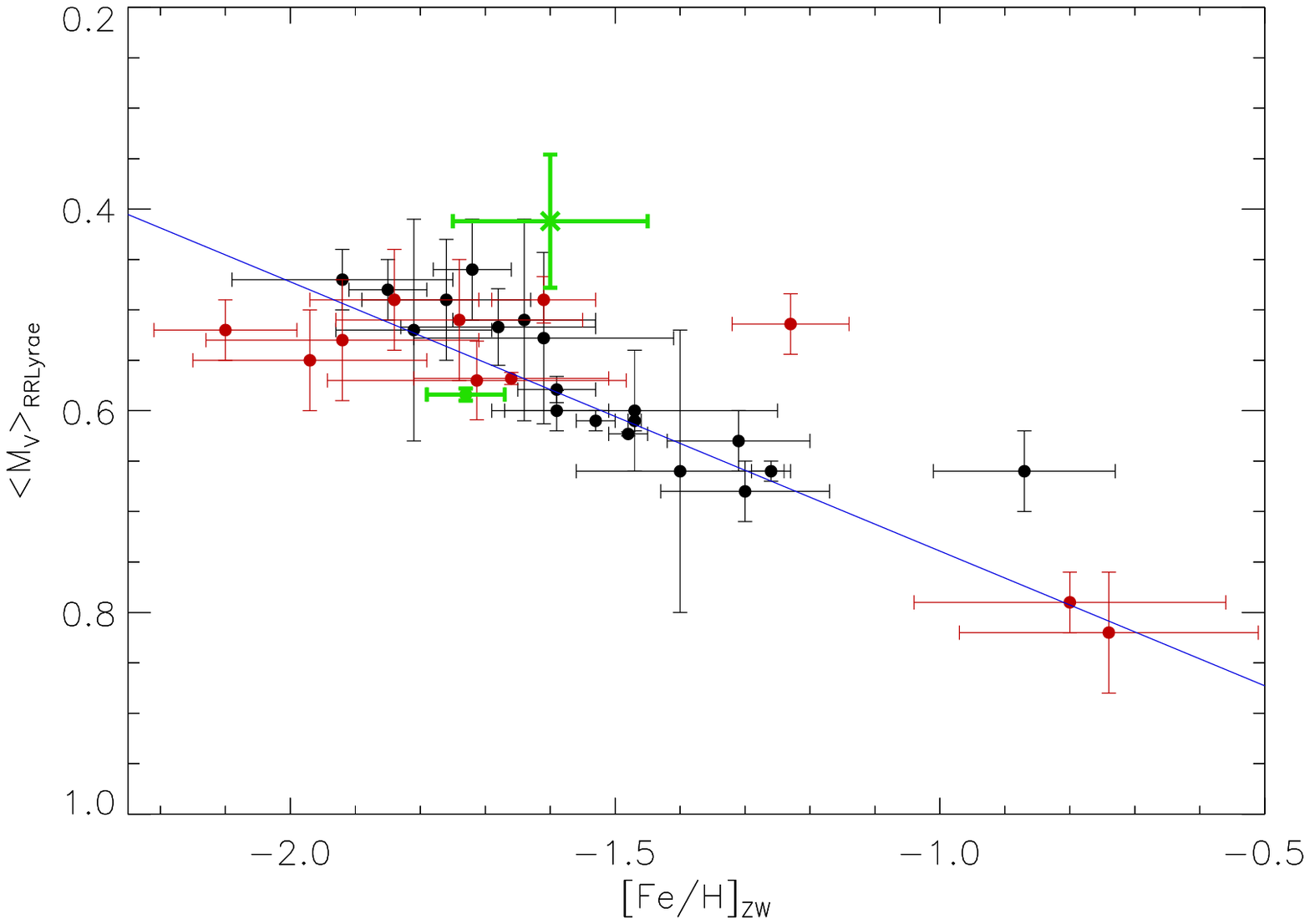}

  \caption{$<M_V>$ vs [Fe/H] for clusters in the literature for which Fourier decomposition of the RR Lyrae variables was performed. Points obtained by analysing the light curves of RR0 variables are shown as black filled circles, while points from analysis of RR1 variables are marked as red filled circles; the data for NGC 1904 are shown as a green cross (RR0 variables) and a green filled circle (RR1 variables). The best linear fit is shown as a blue line, with coefficients given in the text. All points are plotted with 1-$\sigma$ error bars. \label{fig:mvfeh}}

\end{figure}

Although recent theoretical models of the horizontal branch (HB) (e.g. \citealt{vandenberg00}) predict that the relation between $M_V$ and [Fe/H] is non-linear \citep[e.g.][]{cassisi99}, it has traditionally been expressed linearly in the literature in the form $M_V = \alpha\, \rm{[Fe/H]} + \beta$. In this section we compare our values of these parameters for NGC 1904 with other clusters for which Fourier decomposition of RR Lyrae was performed. These are plotted in \Fig{fig:mvfeh} and the data are listed in \Tab{tab:lit_mvfeh}. Note that when values in the literature were not given on the ZW scale, we converted them to the ZW scale to ensure a homogeneous sample. Similarly, the values of $M_V$ for each sample were all converted to values on the ``long" distance scale. In several cases, this meant adjusting the value of $M_V$ by subtracting 0.2 to the quoted value mag to bring it from the scale used by \cite{kovacs98} to the scale of \cite{cacciari05}. There were some exceptions, notably for NGC 6388 and NGC 6441, as it is unclear whether the shift suggested by \cite{cacciari05} applies to metal-rich clusters. The scale used in the original published value for each cluster is noted in \Tab{tab:lit_mvfeh}.

Our best linear fit is $M_V = (0.16 \pm 0.01)\,\fehzw + (0.85 \pm 0.02)$, or, using the UVES scale of \cite{carretta09}, $M_V = (0.14 \pm 0.01)\, \fehuves + (0.81 \pm 0.02)$. This is in good agreement with \cite{arellano08b} who found $M_V = (0.18 \pm 0.03)\,\fehzw + (0.85 \pm 0.05)$. It is also in good agreement with the value found by \cite{fusipecci96}, who analysed the CMDs of eight globular clusters in M31 to derive the absolute magnitude of the HB at the instability strip, and found $M_V = (0.13 \pm 0.07)\,\fehzw + (0.95 \pm 0.09)$, although \cite{federici12} recently found $M_V = (0.25 \pm 0.02)\,\fehzw + (0.89 \pm 0.03)$ for the M31 clusters, a steeper slope than our value. We can also compare our values to some of the many estimates of the slope $\alpha$ and the intercept $\beta$ in the literature, e.g. $(\alpha, \beta)=$(0.22, 0.89) \citep{gratton03}, (0.214, 0.88) \citep{clementini03}, (0.18, 0.90) \citep{carretta00}, all in excellent agreement with our derived value.

The point corresponding RR0 stars on \Fig{fig:mvfeh} appears to be an outlier in the distribution of $M_V$ vs. [Fe/H]. Although this could be the result of poor light curve decomposition, we verified in \Sec{sec:fourdec} that the number of harmonics we fit has little influence on the parameters we derive for the RR Lyrae stars, and therefore on the cluster parameters, so this is unlikely to be the reason for the position of that data point on \Fig{fig:mvfeh}. Since the sample size is small, and error bars are quite large, we refrain from claiming that NGC 1904 is a clear outlier of the distribution, being only $\sim 2 \sigma$ away from our best linear fit of $M_V$ vs. [Fe/H].

% =====================================================
\begin{table*}
\begin{center}
  \begin{tabular}{cccccccc}

     \hline
    NGC \#		& Messier \#	&$M_V$		&$\fehzw$ 	&$M_V$	  &$\fehzw$ 	& Reference	&Distance\\
    			&			&(RR0)		& (RR0)	 	&(RR1)	  &(RR1)		& 			&scale\\

 \hline
NGC 1904 &M 79 	&	$0.41 \pm 0.07$ & $-1.60 \pm  0.15$ & $0.58 \pm 0.01$    & $-1.73 \pm 0.06$  & This work 	& $a$\\
NGC 6981 &M 72	& $0.62 \pm 0.00$ & $-1.48 \pm  0.03$ & $0.57 \pm 0.01$ & $-1.66 \pm  0.15$ & \cite{bramich11} 	& $a$\\
%NGC 1466 & $0.58 \pm 0.01$ & $-1.59 \pm  0.06$ & $0.49 \pm 0.02$ & $-1.61 \pm  0.08$ & \cite{kuehn11} 		& $b$\\
NGC 5024 &M 53	& $0.46 \pm 0.05$ & $-1.72 \pm  0.06$ & $0.49 \pm 0.05$ & $-1.84 \pm  0.13$ & \cite{arellano11} 	& $a$\\
NGC 5286 &$-$	& $0.52 \pm 0.04$ & $-1.68 \pm  0.15$ & $0.57 \pm 0.04$ & $-1.71 \pm  0.23$ & \cite{zorotovic10} 	& $b$\\
NGC 5053 &$-$	& $0.49 \pm 0.06$ & $-1.76 \pm  0.13$ & $0.55 \pm 0.05$ & $-1.97 \pm  0.18$ & \cite{arellano10} 	& $a$\\
NGC 6266 &M 62	& $0.63 \pm 0.03$ & $-1.31 \pm  0.11^*$ & $0.51 \pm 0.03$ & $-1.23 \pm  0.09$ & \cite{contreras10}	& $b$\\
NGC 5466 &$-$	& $0.52 \pm 0.11$ & $-1.81 \pm  0.12$ & $0.53 \pm 0.06$ & $-1.92 \pm  0.21$ & \cite{arellano08b}	& $a$\\
NGC 6366 &$-$	&  $0.66 \pm 0.04$ & $-0.87 \pm  0.14$ & $-$     & $-$  & \cite{arellano08a} 					& $c$\\
NGC 7089 &M 2 	& $0.51 \pm 0.10$ & $-1.64 \pm  0.11^*$ & $0.51 \pm 0.06$ & $-1.74 \pm  0.19^*$ & \cite{lazaro06} 	& $b$\\
NGC 7078 &M 15 	& $0.47 \pm 0.03$ & $-1.92 \pm  0.17^*$ & $0.52 \pm 0.03$ & $-2.10 \pm  0.11^*$ & \cite{arellano06} 	& $b$\\
NGC 5272 &M 3 	& $0.60 \pm 0.02$ & $-1.59 \pm  0.08^*$ & $0.57 \pm 0.04$ & $-$ & \cite{cacciari05} 				& $a, d$\\
NGC 4147 &$-$	& $0.60 \pm 0.06$ & $-1.47 \pm  0.22^*$ & $0.52 \pm 0.07$ & $-$ & \cite{arellano04} 				& $b$\\
NGC 6388 &$-$	& $0.66 \pm 0.14$ & $-1.40 \pm  0.16$ & $0.82 \pm 0.06$ & $-0.74 \pm  0.23$ & \cite{pritzl02} 	& $b^1$\\
NGC 6441 &$-$	& $0.68 \pm 0.03$ & $-1.30 \pm  0.13$ & $0.79 \pm 0.03$ & $-0.80 \pm  0.24$ & \cite{pritzl01} 	& $b^1$\\
NGC 6362 &$-$	& $0.66 \pm 0.01$ & $-1.26 \pm  0.03^*$ & $-$     & $-$  & \cite{olech01} 						& $b$\\
NGC 6934 &$-$	& $0.61 \pm 0.01$ & $-1.53 \pm  0.04$ & $-$     & $-$  & \cite{kaluzny01} 			& $b$\\
NGC 5904 &M 5 & $0.61 \pm 0.01$ & $-1.47 \pm  0.01^*$ & $-$  & $-$  & \cite{kaluzny00} 				& $b$\\
NGC 6333 &M 9 & $0.48 \pm 0.03$ & $-1.85 \pm  0.06^*$ & $-$  & $-$  & \cite{clement99} 				& $b$\\
NGC 6809 &M 55 & $0.53 \pm 0.09$ & $-1.61 \pm  0.20^*$ & $-$ & $-$  & \cite{olech99} 				& $b$\\

	\hline \hline
  \end{tabular}

  \caption{$M_V$ and $\fehzw$ estimates for other clusters in the literature for which these parameters were calculated through Fourier decomposition of the light curves of RR Lyrae stars. All values of $M_V$ were converted to a value consistent with the distance scale of \cite{cacciari05} if another scale was used. Distance scales used in the original publication are: $a$ \cite{cacciari05}; $b$ \cite{kovacs98}; $c$ \cite{cacciari05} but with a shift of 0.18 mag rather than 0.2 mag compared to \cite{kovacs98}; $d$ \cite{cacciari05} but with a shift of 0.23 mag rather than 0.2 mag compared to \cite{kovacs98}. $^1$The values of $M_V$ for these two clusters were not shifted because of uncertainty as to whether the shift given by \cite{cacciari05} applies to metal-rich clusters. $^*$those metallicity values were published on the scale of \cite{jurcsik95}; values listed here are on the ZW scale and were converted using \Eq{eq:zw}. \label{tab:lit_mvfeh}}
  \end{center}
\end{table*}
% =====================================================

% =====================================================
% =====================================================

\section{The distribution of variables in NGC 1904}\label{sec:vardist}

\begin{figure}
  \centering
  \includegraphics[width=8cm, angle=0]{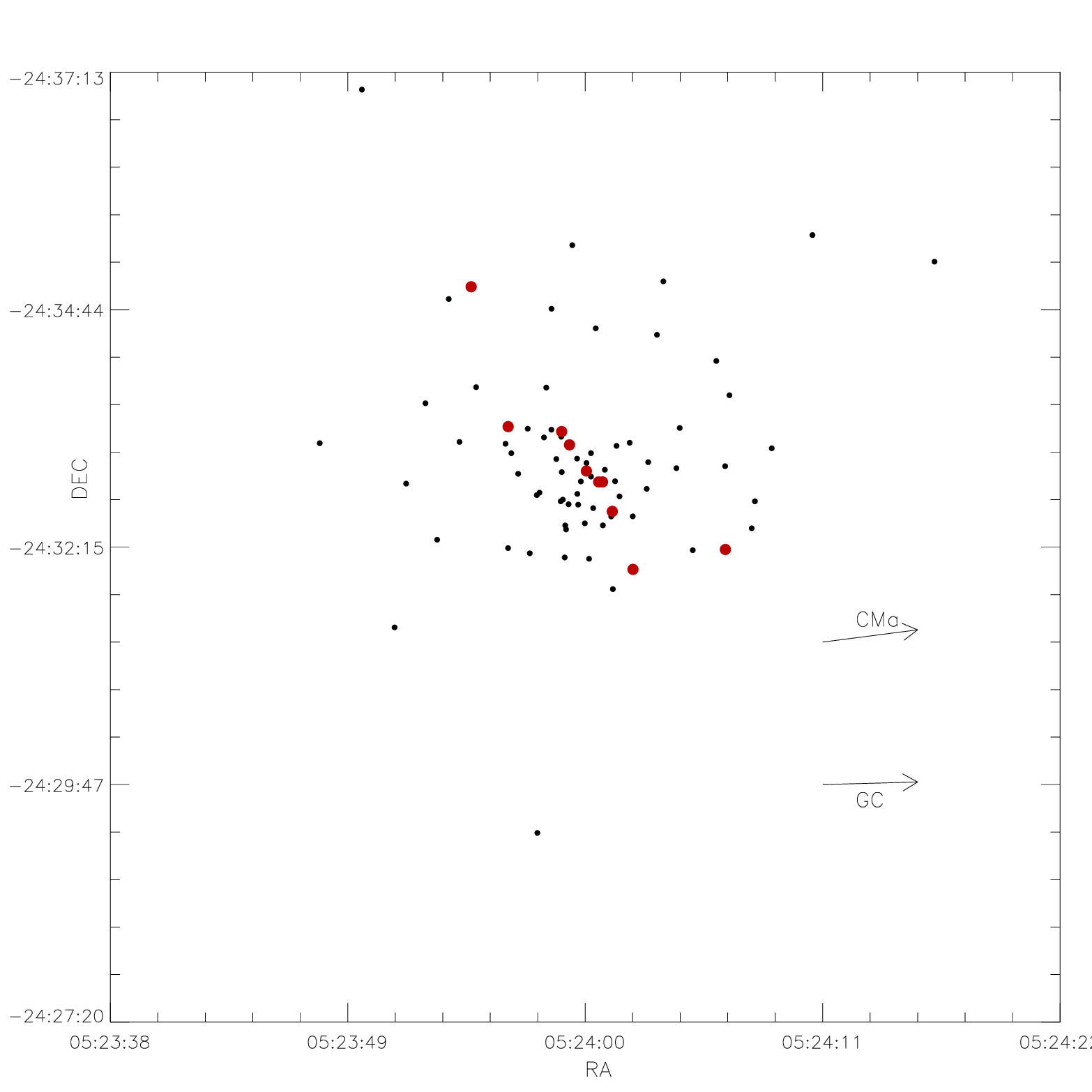}

  \caption{Plot of the spatial distribution of all HB stars (black filled circles) and RR Lyrae stars (red filled circles) on our images. North is up and East to the right. The directions to the Galactic Centre (GC) and Canis Major dwarf galaxy (CMa) are also shown. The tidal radius $r_t\sim 500''$ \citep{lanzoni07} lies outside our plot. \label{fig:hb_dist}}

\end{figure}

From the finding chart of variable stars in NGC 1904 on \Fig{fig:fchart}, we can see that the variables in this cluster are distributed peculiarly, with the variables lying along the Northwest - Southeast axis, rather than distributed randomly across the cluster as would be expected. In order to assess the statistical significance of this, we first compared the distribution of the RR Lyrae variables we detect to the distribution of HB stars in the cluster. This is shown in \Fig{fig:hb_dist}, and the distribution of HB stars can be considered to be spherically symmetric, unlike the distribution of variable stars.

\begin{figure}
  \centering
  \includegraphics[width=8cm, angle=0]{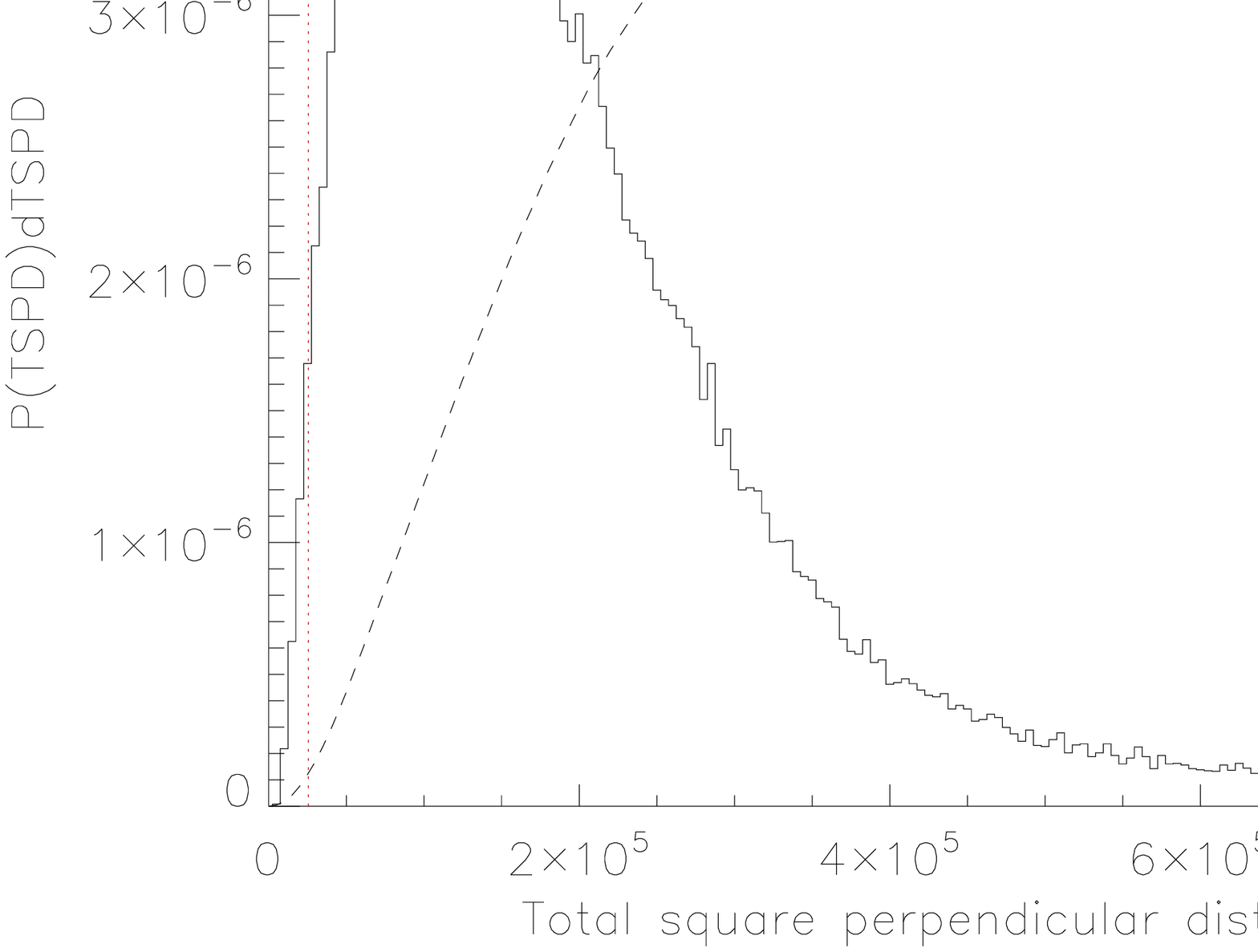}

  \caption{Histogram distribution of the total square perpendicular distance (TSPD) for a linear fit to $10^5$ random samples of 10 HB stars, compared to $\rm TSPD_{var}$, the TSPD of the fit to the RR Lyrae stars in NGC 1904 (red vertical dotted line). The cumulative probability distribution is also plotted as a dashed line; the probability that the value of TSPD is equal to or lower than $\rm TSPD_{var}$ is $\sim 2.8\%$. \label{fig:prob_hb}} \end{figure}

In order to assess the significance of this, we drew $10^5$ random samples of 10 stars from the HB star population in our data and fitted a linear function $y=ax+b$ to each of these, minimising the total square perpendicular distance (TSPD) as a measure of how aligned the stars in the sample are. The TSPD is calculated as

\begin{equation}\label{eq:spd}
{\rm TSPD} = \sum_{i=1}^{N} \left(\frac{| a x_i - y_i + b|}{\sqrt{a^2 + 1}}\right)^2 \, ,
\end{equation}

\noindent
where $N=10$ is the number of variables in the sample. We then compared this to the same statistic for the fit to the RR Lyrae stars in the cluster, $\rm TSPD_{var}$. We find that the probability that a random draw from the overall distribution is equal to or lower than $\rm TSPD_{var}$ is $\sim 2.8\%$. The resulting distribution of TSPD is shown on \Fig{fig:prob_hb}.

% =====================================================
% =====================================================

% =====================================================
% =====================================================

%% =====================================================
%\begin{table*}
%\begin{center}
%  \begin{tabular}{cccc}
%\hline
%Star	&$V$ position	&$I$ position\\
%\hline
%    V1		&1072.860, 1499		&887.599, 1692	\\	%rms=0.015655
%    V2		&saturated 						\\
%    V3		&1100.583, 1046		&915.594, 1238	\\
%    V4		&1295.177, 1004		&1110.963, 1195\\
%    V5		&950.427, 756			&764.899, 947\\
%    V6		&759.807, 451			&573.189, 640\\
%    V7		&1062.334, 888		&877.189, 1078\\		% rms=0.22269
%    V8		&saturated		\\
%    V9		&1057.031, 924		&871.895, 1114	\\
%    V10	&1036.471, 862		&851.294, 1052 	\\
%    V11	& 1028.444, 862		&843.087, 1053	\\		
%    V12	&1002.535, 839.692		&817.912, 1028	\\		
%    V13	&966.909, 784			&781.358, 974	\\
%    V14	&837.743, 746			&651.561, 935	\\	
%	\hline \hline
%  \end{tabular}
%
%  \caption{For own use only: star positions on images in V and I. \label{tab:var_xy}}
%  \end{center}
%\end{table*}
%% =====================================================

\section{Conclusions}

Using difference image analysis, we were able to obtain good photometry for the stars in NGC 1904, even in the crowded central region, where other methods struggle to overcome problems caused by the crowded field. Although the photometry we have obtained for this cluster is not as accurate as that which we obtained for other clusters in previous studies \citep[e.g.][]{arellano10, bramich11}, we discovered a new RR1 variable and verified that one object is not in fact variable. Furthermore, the long time baseline of almost 8 years allowed us to derive precise periods for the RR Lyrae. This in turn has enabled us to estimate some of the properties of the cluster through Fourier decomposition of the RR Lyrae light curves. 

Using this we found a metallicity for NGC 1904 of $\fehzw=-1.63 \pm 0.14$, or, on the scale of \cite{carretta09}, $\fehuves =-1.57 \pm 0.18$. We also find distance moduli of $\mu_0 = 15.64 \pm 0.07$ and $\mu_0 = 15.54$ for RR0 and RR1 variables, translating into distances of $13.26 \pm 0.41$ kpc (using RR0 variables) or 12.87 kpc (using the one RR1 variable with a good period estimate).

Finally, we also used our CMD to check that the metallicity and distance modulus we derived for NGC 1904 is broadly consistent with the theoretical isochrones of \cite{vandenberg06}, and found best-fit isochrones in agreement with the spread of ages reported in the literature.

\section*{Acknowledgements}
NK acknowledges an ESO Fellowship. The research leading to these results has received funding from the European Community's Seventh Framework Programme 
(/FP7/2007-2013/) under grant agreement No 229517. AAF acknowledged the support of DGAPA-UNAM through project IN104612. AAF and SG are 
thankful to the CONACyT (M\'exico) and the Department of Science and Technology (India) for financial support under the Indo-Mexican collaborative project DST/INT/MEXICO/RP001/2008. We thank the referee for constructive comments.

\bibliographystyle{aa}
\bibliography{../thesisbib}

\label{lastpage}

\end{document}